\documentstyle[preprint,eqsecnum,aps,epsf]{revtex}	

\newif\iftightenlines\tightenlinesfalse
\tightenlines\tightenlinestrue

\begin{document}
%
\def\eslt{E\llap/_T}
\def\esl{E\llap/}
\def\msl{m\llap/}
\def\psl{p\llap/}
\def\to{\rightarrow}
\def\te{\tilde e}
\def\tmu{\tilde\mu}
\def\ttau{\tilde\tau}
\def\tl{\tilde\ell}
\def\ttau{\tilde \tau}
\def\tg{\tilde g}
\def\tnu{\tilde\nu}
\def\tell{\tilde\ell}
\def\tq{\tilde q}
\def\tb{\tilde b}
\def\tst{\tilde t}
\def\tt{\tilde t}
\def\tw{\widetilde W}
\def\tz{\widetilde Z}

\hyphenation{mssm}
%
\preprint{\vbox{\baselineskip=14pt%
   \rightline{FSU-HEP-960601}\break 
   \rightline{UH-511-850-96}
}}
\title{SUPERSYMMETRY STUDIES AT \\
FUTURE LINEAR $e^+e^-$ COLLIDERS}
\author{Howard Baer$^1$, Ray Munroe$^1$ and Xerxes Tata$^2$}
\address{
$^1$Department of Physics,
Florida State University,
Tallahassee, FL 32306 USA
}
\address{
$^2$Department of Physics and Astronomy,
University of Hawaii,
Honolulu, HI 96822 USA
}
\date{\today}
\maketitle

\begin{abstract}

We examine various aspects of supersymmetric particle production
at linear $e^+e^-$ colliders operating at a centre of mass energy of
$\sqrt{s}=500$ GeV, 
and integrated luminosity of $\int {\cal L}dt=20-50$ fb$^{-1}$.
Working within the framework of the minimal supergravity model with
gauge coupling unification and radiative electroweak symmetry breaking
(SUGRA), we study various signatures for detection of
sparticles, taking into account their cascade decays,
and map out the regions of parameter 
space where these are observable. We also examine
strategies to isolate different SUSY processes from another. 
In addition, we perform four detailed SUGRA case studies 
and examine the detectability of sparticles when several SUSY
processes are simultaneously occuring. We show that precision
mass measurements of neutralinos, sneutrinos and top-squarks are possible,
in addition to previously studied
precision mass measurements of sleptons and charginos.

\end{abstract}

\medskip
\pacs{PACS numbers: 14.80.Ly, 13.85.Qk, 11.30.Pb}



\section{Introduction}

The realization that weak scale supersymmetry (SUSY) can stabilize
the symmetry breaking sector of the Standard Model (SM) has
made the search for supersymmetric particles\cite{mssm,dpf} one of the
standard items for
experiments at high energy colliders.
Experiments at the CERN LEP2 $e^+e^-$ collider
will soon probe the existence of charginos, sleptons, squarks
and even the lightest of the Higgs scalars of supersymmetric models
if they are lighter than about $\simeq 80-95$ GeV\cite{lep2}.
Direct searches for gluinos and
squarks lighter than about 300~GeV will be carried out at
the Fermilab Tevatron Main Injector
$p\bar p$ collider, which should commence operation
towards the end of the century. Assuming the unification
of gaugino masses, experiments at the Main Injector
may have their
greatest reach for supersymmetry (SUSY) via the clean trilepton channel from
$\tw_1\tz_2\to 3\ell$ production, and ought to be able to probe chargino masses
ranging from $\sim 50-200$ GeV (corresponding to gluinos as heavy as
500-600~GeV), depending on the values of model
parameters\cite{fnal}. The supersymmetry reach of Tevatron experiments
is sensitive to the assumption of $R$-parity conservation and may be
significantly larger (smaller) if the lightest supersymmetric particle (LSP)
decays via $e$ or $\mu$ (baryon number) violating interactions\cite{dpf}. 
It thus
appears that while
LEP2 or the Main Injector could well find a SUSY signal, a decisive
exploration of the existence of weak scale supersymmetry would require
a direct investigation of the TeV scale. This is possible, for instance, at
the CERN Large Hadron Collider (LHC)
which can probe gluino and squark masses up to $\sim 1300-2000$
GeV with just 10 fb$^{-1}$ of integrated luminosity\cite{atlas,lhc}.
We will see later that electron-positron colliders operating at a centre of
mass energy of $\sim 1.5$~TeV would have a similar reach for SUSY discovery
as the LHC: a SUSY signal may, of course, be discovered at a collider
operating (much sooner) with a lower value of $\sqrt{s}$.

If signals for
New Physics are first found, either at the LHC or at the
first phase of an $e^+e^-$ collider operating at 300-500~GeV,
the immediate task at hand would be to establish their origin. The strategy
for this
would be quite different at the two facilities. Focussing on supersymmetry
as the origin of these signals, at the LHC, we would expect signals in
various event topologies ($n$-jets plus $m$ leptons plus $\eslt$, with and
without $b$ tags) from
several sparticle production processes. It is just this complex plethora
of signals that would point to their SUSY origin.
In contrast, it is unlikely 
that (except for anticipated degeneracies, {\it e.g.}
for various flavours of sleptons)
several SUSY reaction thresholds will be crossed at the relatively
low centre of mass energy of the first phase of the Linear Collider, so
that signals may be present from just a single source, 
and in just a few (perhaps, only one) channels. This would, of course,
greatly facilitate their interpretation; the price {\it vis-a-vis} the LHC
is that only a few sparticles might be kinematically accessible.
Having found indications for supersymmetry, the next step would be to 
sort out 
the super-particle masses and quantum numbers. The sparticle mass spectrum
could provide clues\cite{peskin} about the mechanism of supersymmetry 
breaking, and perhaps also provide a window to the symmetries of  
GUT or Planck scale physics. The detailed measurement of sparticle
properties may well be very difficult
at a hadron collider, owing to, among other things, large backgrounds, 
indefinite subprocess collision energy, additional QCD radiation and 
the fact that, in general, several 
SUSY reactions may simultaneously contribute to any one signal.

Many groups have been exploring the physics capabilities of a new, high
energy linear $e^+e^-$ collider\cite{nlc1,nlc,nlc3} 
where the cleanliness of the
experimental environment makes it relatively easy to detect signals from
the production and subsequent decays of new, heavy particles if
sufficient luminosity is available. We will refer to this machine as
the Next Linear
Collider, or NLC. In its
first stage the NLC would operate at a centre of mass energy of
$\sqrt{s}\simeq 300-500$ GeV, and accumulate about 20-50 fb$^{-1}$ of 
integrated luminosity over the first several years of operation. 
It is envisioned that the energy of the machine would be upgraded in stages to 
the TeV region. In addition,
there exists the possibility of longitudinal 
electron beam polarization, perhaps reaching
magnitudes of $\sim 90-95\%$ left- or right- polarization. 

The SUSY discovery
potential of experiments at such a facility has been the subject of
many previous studies. In Ref. \cite{grivaz}, it
has been shown that using relatively simple cuts, $\tw_1\overline{\tw_1}$ 
can be
detected above backgrounds over almost the entire kinematically
accessible regions of parameter space by searching for mixed leptonic/hadronic
or purely hadronic decays of the chargino pair. In Ref. \cite{becker}, it has
been shown that sleptons can usually be detected above background
for slepton masses up to $\sim 90\%$ of the beam energy. Ref. \cite{feng}
shows that with the availability of beam polarization, it should
be possible to measure squark masses to about $\sim 5$~GeV, even taking
their cascade decays into account: in particular, it is possible to obtain
the difference between the masses of $\tq_L$ and $\tq_R$.
Finally, in a pioneering paper, two very detailed case studies 
have been performed by Tsukamoto {\it et. al.}\cite{jlc}, 
showing that an $e^+e^-$ linear collider operating at 
$\sqrt{s}\sim 300-500$ GeV can make a variety of precision measurements 
(which test the assumptions underlying the supergravity GUT framework) of
sparticle masses, spins and coupling parameters. 
This study makes innovative use of the 
capability for polarization of the electron beam. It is, however, assumed
that the beam energy is adjustable so that it is sufficient to focus
on signals from a single SUSY reaction, taken to be
chargino or slepton pair production --- this simplifies the analysis
in that the parent sparticle decays directly to the LSP; {\it i.e.} there
are no complicated cascades to be untangled, and further,
there is no need to sort out various SUSY processes from one
another. The purpose of this paper is to expand upon these studies in
order to attain
two broad goals: 1.) to delineate the reach of such a machine 
for supersymmetry and compare
it to the reach of the LHC, and 
2.) to ascertain to what extent precision measurements of sparticle 
masses are possible, even when several SUSY
processes are occuring simultaneously, and cascade decays of sparticles
are operative.

We work within the
framework of the minimal supergravity GUT model (SUGRA) as defined
in Ref.\cite{bcmpt}. The only role of ``supergravity'' is to provide
a rationale for the universal boundary conditions at an ultra-high
scale $M_X$ which we take to be $M_{GUT}$.
This model is then completely
specified\cite{mssm} by four
SUSY parameters (in addition to SM masses and couplings).
A hybrid set consisting of the common mass $m_0$ ($m_{1/2}$) for all scalars
(gauginos), a common SUSY-breaking trilinear coupling $A_0$ all specified
at the scale $M_X$ together with $\tan\beta$ proves to be a convenient choice.
These parameters fix the masses and couplings of all sparticles. In
particular, $m_{A}$ and the magnitude (but not the sign) of $\mu$ are fixed.

The SUGRA framework (and also a SUGRA-inspired MSSM framework
without radiative elecroweak symmetry breaking\cite{rad})
has been incorporated\cite{bcmpt} into the event
generator program ISAJET 7.16\cite{isajet}.
All lowest order $2\to 2$ sparticle and Higgs boson
production mechanisms have been incorporated into ISAJET. These include
the following processes\cite{bbkmt} (neglecting bars over anti-particles):
\begin{eqnarray*}
e^+e^-&\to & \tq_L\tq_L,\ \tq_R\tq_R ,\\
e^+e^-&\to & \tl_L\tl_L,\ \tl_R\tl_R,\ \te_L\te_R ,\\
e^+e^-&\to & \tnu_{\ell}\tnu_{\ell},\\
e^+e^-&\to & \tw_1\tw_1,\ \tw_2\tw_2,\ \tw_1\tw_2 ,\\
e^+e^-&\to & \tz_i\tz_j,\ (i,j=1-4),\\
e^+e^-&\to & Z H_{\ell},\ Z H_h,\ H_p H_{\ell},\ H_p H_h,\ H^+ H^-.
\end{eqnarray*}
In the above, $\ell =e,\ \mu$ or $\tau$. All squarks (and also all sleptons
other than stau)
are taken to be $L$ or $R$ eigenstates except the stops, 
for which $\tst_1\tst_1$,
$\tst_1\tst_2$ and $\tst_2\tst_2$ (here, $\tst_{1,2}$ being the lighter/heavier
of the top squark mass eigenstates) production is included.
Given a point in SUGRA space, and a collider energy, ISAJET generates all
allowed production processes according to their relative cross sections.
The produced sparticles or Higgs bosons are then decayed into all
kinematically accessible
channels, with branching fractions calculated within ISAJET.
The sparticle decay cascade terminates with the (stable)
lightest SUSY particle (LSP), taken to be the lightest neutralino ($\tz_1$).
Final state QCD radiation is included, as well as particle hadronization.
ISAJET currently
neglects spin correlations, sparticle decay
matrix elements, and also, initial state photon radiation.
In the above reactions, spin correlation effects are only important
for chargino and neutralino pair production, while decay matrix elements 
are only important for three-body sparticle decays. 

To facilitate investigation
of polarized beam effects on signal and background cross 
sections\cite{polpaps}, we 
have recently incorporated polarized beam effects into the ISAJET $e^+e^-$
cross sections. The degree of longitudinal beam polarization has been 
parametrized as
\begin{eqnarray*}
P_L(e^-)&=&f_L-f_R,\ \ \ {\rm where}\\ 
f_L&=&{n_L\over n_L+n_R}={{1+P_L}\over 2}\ \ \ {\rm and}\\
f_R&=&{n_R\over n_L+n_R}={{1-P_L}\over 2}.
\end{eqnarray*}
In the above, $n_{L,R}$ are the number of left (right) polarized 
electrons in the beam, 
and $f_{L,R}$ is the corresponding fraction. Thus, a 90\% right polarized 
beam would correspond to $P_L(e^-)=-0.8$, and a completely unpolarized beam
corresponds to $P_L(e^-)=0$. 

In Sec.~II of this paper, we make a rough scan of 
parameter space
to delineate the regions where a signal for supersymmetry may be observed
above background. We also
delineate those regions where signals from different production processes
can be sorted out from one other. 
A complete visual display of the 4+1 dimensional 
parameter space is of course not possible. Instead, we present results in the
$m_0\ vs.\ m_{1/2}$ parameter plane, for a low and intermediate 
value of $\tan\beta$
($\tan\beta =2$ and 10), and for both $\mu <0$ and $\mu >0$. Variation of 
the parameter $A_0$ mainly leads to changes in third generation squark
and slepton masses, especially for the top squarks $\tst_1$ and $\tst_2$.
Hence, we do not address in detail consequences of $A_0$ variation, 
although in Sec. 3 we perform a case study where the $\tst_1$ becomes
light enough to be accessible. 
A recent analysis of production 
and decay rates of third generation sparticles at linear $e^+e^-$ colliders
can be found in Ref. \cite{newbartl}.
In most of the work presented here, we 
set $A_0=0$.

Our objective in Sec.~III can be viewed as a follow-up to reference\cite{jlc}.
Here, we perform a number of case studies in some detail to explore the
capability of an NLC to perform precision measurements of sparticle
masses and parameters. We allow for the simultaneous production
of all sparticles with cross sections and decay patterns as given
by the model, and describe our attempts to extract the masses of
various sparticles.
We examine the following four cases, which have
also been the subject of recent studies on supersymmetry at various
workshops on linear colliders\cite{workshops}. 
For each case, we list 
($m_0,\ m_{1/2},\ A_0,\ \tan\beta\ {\rm and}\ sign(\mu )$), with the
mass parameters in GeV. 
We have taken $m_t=180$ GeV throughout this paper.

\begin{enumerate}
\item $(400,\ 200,\ 0,\ 2,\ -1)$  (dominantly chargino production),
\item $(100,\ 300,\ 0,\ 2,\ -1)$  (dominantly slepton production),
\item $(200,\ 100,\ 0,\ 2,\ -1)$  (mixed chargino/slepton/sneutrino production),
\item $(300,\ 150,\ -600,\ 2,\ +1)$ (includes $\tst_1\bar{\tst_1}$ production),
\end{enumerate}
For each case, we use ISAJET 7.16 to simultaneously produce all 
allowed SUSY particles as well as SM backgrounds. 

We conclude in Sec.~IV with a summary of our results and some comparisons
between the NLC and the CERN LHC $pp$ collider.
Finally, in an appendix, we collect various expressions for 
lowest order production cross sections of SM and SUSY particles 
via polarized beams.

\section{Reach of the NLC in SUGRA parameter space}

\subsection{Kinematic Reach}

To gain some orientation for our study of
the SUSY reach of the NLC in the parameter
space of the minimal SUGRA model, we begin by examining
where in parameter space the various sparticle production mechanisms
are kinematically accessible.
The lightest SUSY Higgs boson $H_\ell$ is very special, since in the
minimal model, $m_{H_{\ell}}\alt 120-130$ GeV\cite{lighthiggs}, and
so, is accessible either via $e^+e^- \to Z H_\ell$ or 
via $e^+e^- \to H_p H_\ell$ processes for all values of parameters.
Furthermore, if $|\mu|$ is large as is often the case in the SUGRA framework,
its couplings to SM fermions and gauge bosons are expected to be very
similar to those of a SM Higgs boson. Recently, detailed studies of the
detectability
of SM Higgs bosons as well as of the Higgs bosons of the minimal SUSY model
at the NLC have been performed by Janot\cite{janot}.
Even for the most difficult case, where $m_{H_{\ell}}\sim M_Z$, a $5\sigma$
signal should be attainable within a month of running at or near design
luminosity at a 500 GeV linear collider, assuming a $b$ tagging
efficiency of 50\% with a rejection of 97.5\% (99.9\%) against $c\bar{c}$
(light quark pairs) is achieved. Detection of $H_\ell$ should be
possible even if it decays invisibly via $H_\ell \to \tz_1\tz_1$ pairs.
Thus, even if the lightest SUSY Higgs boson
$H_\ell$ eludes detection at LEP2, the Tevatron and the 
LHC, it should certainly be
discovered at NLC. Hence, the NLC will be able 
to exclude the minimal SUGRA model if no $H_{\ell}$ signal is seen\cite{FN1}. 
However, if
an $H_{\ell}$ signal is the only new signal seen, it may well be 
difficult\cite{haber} 
to distinguish whether one has seen a SUSY or a SM Higgs boson:
thus, detection of $H_{\ell}$ alone will likely not be definitive evidence for
supersymmetry. In the remainder of this paper, we assume the lightest 
SUSY Higgs boson $H_{\ell}$ will be detected at least at the NLC, and so focus
our attention on the detectability of 
super-partners.

In Fig. 1, we show the regions of the $m_0\ vs.\ m_{1/2}$
parameter plane where various $2 \to 2$ SUSY particle processes are
kinematically accessible
to an $e^+e^-$ linear collider operating at $\sqrt{s}=500$ GeV, for
$\tan\beta =2$, $A_0=0$ and {\it a}) $\mu <0$ and {\it b}) $\mu>0$. 
In Fig. 2, we show
the corresponding results for $\tan\beta =10$.
The regions labelled by TH are excluded by various
theoretical constraints\cite{bcmpt}, while regions labelled EX are excluded by
experimental searches for SUSY at the LEP\cite{PDG} and Fermilab Tevatron
colliders as described in Ref.\cite{bcmpt} (we do not include constraints from
the LEP1.5 run).
The regions below the contours labelled by a sparticle pair is
where the corresponding reaction is
kinematically accessible. The neutralino pair
contours are an exception. Since the neutralino production
cross section can be strongly suppressed by mixing angle factors,
we conservatively show the regions where the production cross section
$\sigma (\tz_i\tz_j)> 10$ fb ($i=1-4, j=2-4$).  We see
from  Fig. 1{\it a} and {\it b} that the
outermost boundary of the kinematic
reach for SUSY is comprised of three contours: the $\te_R\te_R$
contour for low $m_0$, the $\tw_1\tw_1$ contour for large $m_0$, and a small
intermediate region where the $\tz_1\tz_2$ reaction might be accessible.
The situation is similar in Fig. 2, except that an additional sliver of
parameter space may be accessible by searching for $\ttau_1\ttau_1$ pairs
(because of $\tau_L-\tau_R$ mixing induced by the tau Yukawa coupling
which increases with $\tan\beta$, the
lighter of the two staus is lighter than $\te_R$)
as well; in practice, the detection efficiency is larger\cite{lep2}
for identifying
selectron and smuon signals so the $\tau$ channel 
is unlikely to be relevant for
the maximal reach. The smuon contours (as well as stau contours in Fig.~1)
essentially overlap with the selectron contours and have
not been shown. However, the detection and measurement
of their properties is very important for testing slepton
universality\cite{jlc}
and the flavour structure of the slepton sector: this could shed
light\cite{hall}
on the nature of physics at the GUT scale.
Interesting information about the composition (gaugino versus Higgsino) 
of $\tz_1$
\cite{nojiri} can also be 
obtained from a detailed study of the stau signal.
For parameter space points outside these regions, SUSY particles will be 
accessible at NLC500 only via higher order processes\cite{highord},
while inside these regions,
at least one and often many SUSY particles might be produced with significant 
rates. 

In the low $m_0$ region, we see that various slepton pair reactions--
$\tell_R\tell_R$, $\te_R\te_L$, $\tell_L\tell_L$ and $\tnu_\ell\tnu_\ell$--
can be accessible. The sneutrinos may or may not have visible decay
products, depending on how massive they are relative to the charginos and
neutralinos.
Potentially, the greatest reach for $\te_L$ is via the $\te_R\te_L$
production which occurs via $t-$channel neutralino exchange; the production
cross section is, however, significantly smaller\cite{bbkmt} than
$\sigma(\te_R\te_R)$ or $\sigma(\te_L\te_L)$ (if both selectrons
have the same mass). Smuons and staus can only be produced via $s$-channel
processes.
The regions with $m_0\alt 250$ GeV are also the regions 
most favored by cosmological neutralino relic density 
constraints\cite{brhlik}, 
which require
$100\alt m_{\tell}\alt$ 250 GeV to obtain a dark matter
relic density in accord with inflationary cosmological models with 1:2 mixed
hot to cold dark matter. 

In the
large $m_0$ region, sfermions are too heavy so that this part of parameter
space
can best be probed by searching for chargino pairs,
and possibly $\tz_1\tz_2$ or $\tz_2\tz_2$ pairs.
Since $\tz_{1,2}$ is gaugino-like for a wide range of SUGRA parameters,
these neutralino pair cross 
sections depend significantly on $t$ or $u$ channel $\te_L$ or $\te_R$ 
exchange, so the neutralino pair production rate typically decreases with
increasing $m_0$. The large $m_0$ region is difficult to accommodote
cosmologically since the annihilation rate of the (stable) LSPs via 
sfermion exchanges is 
suppressed; this implies too short an age for our universe
unless $\tz_1\tz_1$ annihilation can efficiently proceed via $s$-channel
$Z$ or $H_\ell$ resonances\cite{brhlik}.

In the regions with smaller $m_0$ and $m_{1/2}$, the higher mass chargino
and neutralino states $\tw_2$, $\tz_3$ and $\tz_4$ may become 
accessible to NLC500 searches. 
For still lower $m_0$ and $m_{1/2}$, $\tst_1\tst_1$
production and ultimately $\tq\tq$ production can become accessible. However,
for NLC500, $m_{\tq}$ must be less than 250 GeV to be at all accessible, 
so that these strongly interacting states would have been seen much 
earlier\cite{fnal} 
at Tevatron collider experiments. Finally, we see that in the lowest
$m_0$ and $m_{1/2}$ regions, heavy Higgs bosons such as $H^\pm$ may become
accessible to NLC searches.

By comparing Fig. 1{\it a} and {\it b} with Fig. 2{\it a} and {\it b}, 
we see that although the 
sparticle accessibility contours can change somewhat with variations in 
$\tan\beta$ or $sign(\mu )$, the overall qualitative trends are the same: 
NLC500
is most likely to access the non-colored, charged SUSY particle states
and sneutrinos,
with little hope of seeing squarks or gluinos, unless they have already been 
discovered at Tevatron experiments. The exception is the possibility
of seeing third generation $\tst_1$ or $\tb_1$ squarks, which can have
significantly lighter masses than the other squarks, which should be nearly
mass degenerate in the minimal SUGRA model. Variation of the $A_0$ 
parameter mainly affects third
generation $\tst_i$, $\tb_i$ and $\ttau_i$ masses ($i=1$ or 2), so that only
their reach contours change appreciably with $A_0$.

\subsection{Results from Event Simulation}

We now turn to the evaluation of the reach of the NLC
by comparing signal against SM backgrounds using explicit event
generation. Since beam polarization has been shown to be a useful tool,
as a first step, we show in Fig. 3 various lowest order SM
background
cross sections, as a function of electron polarization $P_L(e^-)$
for an unpolarized positron beam.
The $e^+e^-\to e^+e^-$ contribution includes only $s$-channel contributions.
For unpolarized beams ($P_L(e^-)=0$),  $WW$ production is the dominant SM
process. By tuning $P_L(e^-)$ to $\sim -0.9$ (95\% right polarized beam),
the magnitude of the $WW$ cross section can be significantly reduced
relative to other SM backgrounds, which show only a mild dependence
on beam polarization. Since $WW$ production is a major background for
many new physics processes, Fig.~3 suggests that the use of (dominantly)
right-handed electron beams would yield a better signal to background ratio,
except for those signals whose cross sections become small when
$P_L(e^-) \simeq -1$. We have not shown backgrounds from $2\to3$ and $2\to 4$
SM processes-- these can be reduced by using suitable cuts\cite{jlc}.

Our next step is to generate explicit events for signal and background. We
focus on optimizing cuts for $\te_R\te_R$ and $\tw_1\tw_1$ production,
which should give the largest reach into parameter space.
We use the ISAJET toy detector ISAPLT with the following characteristics.
We simulate calorimetry covering
$-4<\eta <4$ with cell size $\Delta\eta\times\Delta\phi =0.05\times 0.05$.
Energy resolution for electrons, hadrons and muons is taken to be
$\Delta E=\sqrt{.0225E+(.01E)^2}$, $\Delta E=\sqrt{.16E+(.03E)^2}$ and
$\Delta p_T =5\times 10^{-4} p_T^2$ respectively.
Jets are found using fixed cones of size
$R=\sqrt{\Delta\eta^2+\Delta\phi^2} =0.6$ using the ISAJET routine
GETJET (modified for clustering on energy rather than transverse energy).
Clusters with
$E>5$ GeV and $|\eta ({\rm jet})|<2.5$ are labeled as jets.
Muons and electrons are classified as isolated if they have $E>5$ GeV,
$|\eta (\ell )|<2.5$, and the visible activity within a cone of $R=0.5$
about the lepton direction is less than
$max({E_{\ell}\over 10},\ 1\ {\rm GeV})$. Finally, $b$-jets are
tagged with an efficiency of $50\%$, while $c$-jets are
misidentified
as $b$'s with an efficiency of 3\%. Jets with one or three charged
prongs are classified as $\tau$s for the purpose of $\tau$-veto (see
Sec.~IIID).

The signature for $\tell_R\tell_R$ production is a pair of acollinear
same flavor/opposite sign leptons recoiling against $\esl$. To search for
such a signal, we essentially follow the cuts of Ref. \cite{jlc} and require
{\it i}) 5 GeV $<E(\ell )<200$ GeV,
{\it ii}) 20 GeV $<E(visible)<400$ GeV,
{\it iii}) $|m(\ell^+\ell^-)-M_Z|>10$ GeV,
{\it iv}) $|\cos\theta (\ell^{\pm})|<0.9$,
{\it v}) $-Q_{\ell}\cos\theta_{\ell}<0.75$,
{\it vi}) $\theta_{acop.}>30^o$,
{\it vii}) $\eslt >25$ GeV and
{\it viii}) veto events with any jet activity, where the polar
angle is measured from the electron beam, and $Q$ is the charge of the 
lepton. 
Cut {\it iii}) eliminates backgrounds from $e^+e^- \to ZZ,\nu\nu Z$ and
$e^+e^-Z$ production, while cuts {\it iv}) and {\it v}) greatly
reduce the backgrounds
from $WW$ and $e\nu W$ production (we neglect the latter).
For unpolarized beams, the resulting background level was 17 fb, while
for $P_L(e^-)=-0.9$, the background was 2.4 fb.
Thus, for the polarized case, a $5\sigma$ signal for
20 fb$^{-1}$ of integrated luminosity requires a signal rate larger than
1.73 fb.

To search for chargino pairs, one may search either for 4-jet events
from both $\tw_1$ hadronic decays, or 1-$\ell$+2-jet events from
mixed hadronic/leptonic chargino decays. We found that
either signature gives a
similar reach; we focus on the mixed hadronic/leptonic signature since
ultimately it is more useful for chargino mass measurements.
Following Ref. \cite{jlc}, we require events with one lepton plus 2 jets, and
{\it i}) $\#$ of charged tracks $>5$,
{\it ii}) 20 GeV $<E(visible)<400$ GeV,
{\it iii}) if $E(jj)>200$ GeV, then $m(jj)<68$ GeV,
{\it iv}) $\eslt >25$ GeV,
{\it v}) $|m(\ell\nu)-M_W|>10$ GeV for a $W$-pair hypothesis,
{\it vi}) $|\cos\theta (j)|<0.9$, $|\cos\theta (\ell )|<0.9$,
$-Q_{\ell}\cos\theta_{\ell}<0.75$ and $Q_{\ell}\cos\theta (jj)<0.75$
{\it vii}) $\theta_{acop.}(WW)>30^o$ for a $W$-pair hypothesis.
Although the dominant background from $WW$ production is smallest
for $P_L(e^-)$ close to -1, the signal cross section also drops 
rapidly since the chargino is frequently an $SU(2)$ gaugino, and
so, couples only to the doublet electron. Hence, the use
of left-handed electron beams is required.
For $P_L(e^-)=+0.9$, the resultant background level was 155 fb, 
so that a $5\sigma$ signal for
20 fb$^{-1}$ of integrated luminosity requires a signal rate larger than
14 fb.

In Fig. 4 and Fig. 5,
we show the $5\sigma$ reach of NLC500 for minimal SUGRA via
$\ell^+\ell^-$ and 1-$\ell +2$-jets searches for $\tan\beta=2$ and
$\tan\beta=10$, respectively. Here, we assume an integrated
luminosity of 20~fb$^{-1}$ and
compare the reach we obtain with the
kinematic reach contours of Figs.~1 and 2, shown as dashed
contours for the $\ell^+\ell^-$ and $1\ell+2j$ signals.
In Fig.~4{\it a} alone, we compare the reach for a polarized $e^-$ beam
with the unpolarized case.
The dotted curves
correspond to the NLC500 reach using the above cuts with unpolarized beams,
while the solid curves correspond to the reach for a 95\% polarized
electron beam with dominantly left (right) handed polarization for
the chargino ($\tell_R$) search. Notice that for $m_0 \agt 250$~GeV,
the $\ell^+\ell^-$ signal (with unpolarized beams) from {\it charginos}
is observable in between the two dotted curves; below the lowest
dotted curve the chargino is rather light, and our cuts are not
optimised for their selection.
By comparing the dotted and solid curves,
we see that there is only a marginal gain in reach using
polarized beams with
$P_L(e^-)=-0.9$ for the slepton signal
(to reduce $WW$ backgrounds), and $P_L(e^-)=+0.9$ for the chargino signal
(to gain the largest signal cross-section).
For this reason, we have chosen not to show the polarization dependence
in the other frames in Fig.~4 and Fig.~5.
The real power of polarization
is for precision measurements of masses and couplings\cite{jlc,fmpt}.
We note that
the selectron search contours fill
most of the region of slepton accessibility, except for a small region
around $(m_0,m_{1/2})=(100,500)$ GeV, where $m_{\tell_R}\simeq m_{\tz_1}$,
and the two leptons have very little visible energy. The $\ell +$2-jets
signal from chargino pair production can likewise be seen almost up to the
kinematic limit over much of the parameter space; the exception is around
$(m_0,m_{1/2})=(250,275)$ GeV, where the $\tnu_L$ becomes light enough to
cause a significant drop in the chargino pair production total cross section.
For even lower values of $m_0$, the region of detectable charginos falls off
due to rising (diminishing) chargino leptonic (hadronic) branching fractions,
as sleptons become very light, in which case a signal may be observable in
the acollinear $e\mu+\eslt$ channel.

Some additional reach may be gained by looking for $\tz_1\tz_2$
production around $(m_0,m_{1/2})=(250,320)$ GeV. For the $\tan\beta =2$
cases, $B(\tz_2\to\tz_1 H_{\ell})$ is almost 100\%, 
so the $\tz_1\tz_2$ signature
is $b\bar{b}+\esl$. The physics background consists of $ZZ$ and $ZH_{\ell}$
production, which occurs at a much larger rate.
For the $\tan\beta =10$ cases, $\tz_2\to\tz_1 Z$ is comparable to
$\tz_2\to\tz_1 H_{\ell}$, so one
can also look for $Z+\esl$ events, where $Z\to\ell^+\ell^-$.
The $Z+\esl$ signals also have large backgrounds from $WW$ production.
To determine the viability of the $\tz_1\tz_2$ signal in this region, we
examined the point
$(m_0,m_{1/2},A_0,\tan\beta ,sgn(\mu )=(240,340,0,2,+1)$ in parameter space.
In our simulation we used $P_L(e^-)=+0.9$, and assumed 
20 fb$^{-1}$ of integrated luminosity.
To isolate the $\tz_1\tz_2$ signal, we require events
with 2 tagged $b$-jets, $\eslt >25$ GeV and $30^o<\Delta\phi_{b\bar{b}}<150^o$.
We also require that the missing mass $\msl > 340$~GeV since
for this case,
the two main backgrounds ideally have $\msl =M_Z$, up to considerable
smearing corrections; in contrast, the SUSY signal requires $\msl >2m_{\tz_1}$.
For the particular point we considered for our simulation, we found a 
total $\tz_1\tz_2$ cross section of 21 fb, with a signal
efficiency of 6\%. No SUSY or SM background events were found.
We thus conclude that a
left-polarized cross section of $\sigma (\tz_1\tz_2 )\simeq 10$ fb is 
needed to achieve a
$\sim 10$ event signal with 20 fb$^{-1}$ of data. We show in Figs. 4 and 5
the 10~fb contour labelled $\tz_1\tz_2$, 
which represents the rough reach in parameter space for
the $\tz_1\tz_2$ signal, assuming that $\tz_2 \to ZH_{\ell}$ is the dominant
decay of $\tz_2$ and that the detection efficiency varies slowly in this
region of parameter space.

\section{Sparticle Mass Measurements at Linear Colliders:
Four Case Studies}

Our task in this Section is to go beyond the detectability of new
particles, and to examine prospects for the determination of their
properties. Towards this end, we perform
four detailed case studies where we attempt
to isolate the different SUSY
production processes from one another in order to facilitate the
interpretation of our results.
We focus here on precision mass measurements, although certainly 
a wide range of other measurements such as spin 
quantum numbers, total cross sections, sparticle branching 
fractions {\it etc.}
are possible\cite{jlc}. Such measurements can serve as tests of the underlying
framework (the minimal SUGRA model, in our case), and perhaps even help
to determine some of its fundamental parameters.

Since SUSY particle decays always terminate in the LSP $\tz_1$, a direct 
reconstruction of SUSY particle masses 
via ``mass bumps'' is not possible. However, the 
cleanliness of $e^+e^-$ scattering events, combined with the well-defined 
initial state, leads to kinematic restrictions which depend directly on 
sparticle masses. 
For instance, in the reaction
$e^+e^-\to p_1+p_2$, followed by $p_2\to p_3+p_4$, the energy of particle $p_3$
is restricted to lie between
\begin{equation}
\gamma (E_3^*-\beta p_3^*)\le E_3 \le \gamma (E_3^*+\beta p_3^*),
\end{equation}
where $E_3^*=(m_2^2+m_3^2-m_4^2)/2m_2$, 
$p_3^*=\sqrt{E_3^{*2}-m_3^2}$, $\gamma =E_2/m_2$, $\beta=\sqrt{1-1/\gamma^2 }$
and $E_2=(s+m_2^2-m_1^2)/2\sqrt{s}$, up to corrections from energy
mis-measurements, particle losses, bremsstrahlung, etc.
We will see below that this formula provides a simple yet clean
way for the determination of slepton and LSP (or sneutrino
and chargino) masses\cite{jlc} and, with appropriate 
analysis, also of the chargino mass when the chargino decays via $\tw_1 \to
f\bar{f}\tz_1$ (see Sec.~IIID).

\subsection{Case 1: Dominantly chargino production}

The first case that we examine corresponds to the SUGRA point
$(m_0,m_{1/2},A_0,\tan\beta,sgn(\mu ))=(400,200,0,2,-1)$,
(where parameters with mass dimensions are in GeV)
whose locus in the
$m_0\ vs.\ m_{1/2}$ plane is labelled ``1'' in Fig. 4{\it a}.
We show in Fig. 6 the total cross sections for
accessible $2\to 2$ SUSY particle reactions as a
function of beam polarization $P_L(e^-)$.
These may be compared directly to various SM
cross sections shown in Fig. 3. We also plot
the masses of the accessible SUSY particles to help orient the
reader. For this case,
$m_{H_{\ell}}=85$ GeV, so it would likely have been already discovered at LEP2.
The $ZH_{\ell}$ cross section only varies by 50\% over the range of $P_L(e^-)$.
The dominant SUSY process is $\tw_1\tw_1$ production
($m_{\tw_1}=175$ GeV) and, because the chargino is essentially an
$SU(2)$ gaugino, its cross section drops rapidly as $P_L(e^-)\to -1$.
At high energy ($\sqrt{s}>>M_Z$), we may think of just the neutral
$SU(2)$ vector boson exchange contributing in the $s$-channel, so that
$s$-channel amplitudes for right-handed electrons are suppressed;
since the sneutrino exchange amplitude always involves just left-handed
electrons, the polarization dependence of this cross section is
readily understood. The same reasoning explains the behaviour of the
$\tz_2\tz_2$ cross section. In the limit that $\tz_1$ is the bino,
the $t$-channel selectron amplitude dominates $\tz_1\tz_1$ (and $\tz_1\tz_2$)
production. The polarization dependence of $\sigma({\tz_1\tz_1})$ is readily
understood once we recognize that the cross section varies as $Y^4$, where
$Y$ is the hypercharge of the selectron exchanged in the $t$-channel.
Finally, because $\tz_2$ has suppressed hypercharge gauge couplings,
the polarization dependence of $\sigma({\tz_1\tz_2})$ follows that of
$\sigma({\tz_2\tz_2})$. 

In this scenario, $\tw_1\to W\tz_1$ with nearly 100\% branching ratio, so the
$\tw_1\tw_1$ signal should be easily seen above the $5\sigma$ level
of Sec. 2 in either the 4-jet or 1$\ell+$2-jets mode. However, to extract a
chargino mass, a clean event sample is needed, and further discrimination of
signal from SM (mainly $WW$) background is necessary. We focus here on the
1$\ell+$2-jets signal, for which mass measurements are relatively
straightforward. We use unpolarized beams and assume
$\int {\cal L}dt=20$ fb$^{-1}$. The
missing mass, defined by $\msl =\sqrt{\esl^2 -\psl^2}$
provides a powerful discriminator. For $\tw_1\tw_1$
production, $\msl$ is constrained to be $\msl >2m_{\tz_1}=172$ GeV, while
$WW$ production has no such constraint. 
We show in Fig. 7{\it a} the $\msl$
distribution for both signal and background. In this case, a rather clean
SUSY signal can be obtained by requiring $\msl >240$ GeV. The
distribution of surviving
events is plotted as a function of dijet energy $E_{jj}$ in Fig. 7{\it b}.
The background level is indicated by the histogram, while the signal 
cross section is shown by the points with
error bars. In this case, $E_{jj}\simeq E_W$ from the $\tw_1\to W\tz_1$
decay, so the endpoint structure of a 2-body decay discussed
above should apply. The tips of the
arrows indicate the the theoretically expected
endpoints obtained using Eq. (3.1).
The distribution has significant smearing (particularly at the low end)
due to our calorimeter simulation and use of the cone algorithm for
jet finding (which entails some loss from energy outside the cone),
and may well be improved with different jet finding
schemes, or by plotting all visible energy aside from the detected lepton.
Since Tsukamoto {\it et. al.} \cite{jlc}
have already shown
that a fit to the  $E_{jj}$ distribution leads to a
mass measurement of $m_{\tw_1}$ to $\sim 5$\%,
we have not made any attempt to improve our jet algorithm and repeat this same
analysis here.

Instead we focus on the other interesting possibility which
is to isolate a signal from $\tz_2$ in order to
measure its mass. For our case, $B(\tz_2\to\tz_1 H_{\ell})=99.6$\%,
so that $\tz_1\tz_2$ production almost exclusively results in
$b\bar{b}+\esl$ events.
The physics background here is mainly due to $ZZ$ and $ZH_{\ell}$ production.
In this case, we use $P_L(e^-)=+0.9$, and assume 50 fb$^{-1}$ of integrated
luminosity.
To isolate the $\tz_1\tz_2$ signal, we require events
with two tagged $b$-jets, $\eslt >25$ GeV and $30^o<\Delta\phi_{b\bar{b}}<150^o$.
At this point, we can proceed
with a plot of $\msl$, which is shown in Fig. 8{\it a}. For this case,
the two main backgrounds have $\msl =M_Z$, up to considerable
smearing corrections, while the SUSY signal requires $\msl >2m_{\tz_1}=172$ GeV
again. In fact, the kinematic endpoints for $\msl$ in the SUSY case are
easily calculable, since
$\msl =\sqrt{\esl^2-\psl^2}=\sqrt{s-2\sqrt{s}E_{H_\ell}+m_{H_\ell}^2}$
(with $E_{H_\ell}$ bounded as in Eq.~(3.1)), 
and are indicated on the plot. A clean separation between
signal and background can be obtained by requiring $\msl >300$ GeV. 
If the $H_\ell$ mass is well measured from LEP2
or NLC, then these 
endpoints can be used to measure $m_{\tz_2}$ and $m_{\tz_1}$.
The dijet invariant mass from 
$H_{\ell}\to b\bar{b}$ may be checked for consistency with a Higgs boson 
mass hypothesis. We show its distribution in Fig. 8{\it b} 
after the $\msl >300$ GeV cut. It peaks
somewhat below $m_{H_\ell}$ due to imperfect jet reconstruction,
energy mismeasurement and energy loss due to neutrinos. Finally, 
$E_{H_\ell}=E_{b\bar{b}}$ may also be used for a determination of 
$m_{\tz_1}$ and $m_{\tz_2}$. We show the dijet energy 
distribution in Fig. 8{\it c} along with the expected
background 
(solid histogram). The tips of the arrows denote the
theoretically expected end points. Once again, we see that there is
considerable smearing at the lower end. 

The missing mass distribution in Fig.~8{\it a} appears best suited for
a mass measurement because missing energy from mismeasurement or calorimeter
losses cancels out to some degree in constructing $\msl$. We perform a fit
to the $\msl$ distributions, which depends on $m_{\tz_2}$ and $m_{\tz_1}$
(assuming $m_{H_{\ell}}$ is known from LEP2 or NLC), and plot
the resulting $\Delta\chi^2=\chi^2-\chi_{min}^2=2.3$ and 4.6
contours (these correspond to 68\%, {\it i.e.} ``$1\sigma$''
and 90\% CL error ellipses)
for a 50 fb$^{-1}$ ``data'' sample.
The result is shown in Fig.~9{\it a}, from which we see that
$m_{\tz_2}=176.5\pm 4$ GeV, and $m_{\tz_1}=86.1\pm 3$ GeV ($1\sigma$ error
bars), to be compared with 
the input 
values of $m_{\tz_2}=175.2$ GeV and $m_{\tz_1}=85.9$ GeV. The ``data''
(points)
and 
the best fit (histogram) are shown below in Fig.~9{\it b}.

\subsection{Case 2: Dominantly selectron production}

To contrast with Case~1, we now consider a set of parameters for
which selectron pair production dominates, and pair production
of charginos is kinematically forbidden at NLC500.
For Case~2, we choose
$(m_0,m_{1/2},A_0,\tan\beta ,sgn(\mu ))=(100,300,0,2,-1)$.
The cross sections for accessible processes
are shown versus $P_L(e^-)$ in Fig.~10 together with relevant sparticle
masses.
The polarization dependence of the slepton pair production
cross sections is straightforward to understand --- $\tell_R$ pair production
is maximized for a right-handed electron beam because it occurs essentially
due to hypercharge gauge boson exchange, and $Y(\tell_R)=2Y(\tell_L)$ (for
$\te_R\te_R$ production, this effect is further accentuated by
the fact that the $t$-channel exchange amplitude vanishes for $P_L(e^-)=1$).
The production of $\te_L\te_R$ pairs, which occurs only via $t$-channel
exchange, is independent of beam polarization. Finally, the production
of left slepton (or sneutrino) pairs shows the opposite dependence as
$\tell_R$ pair production, but it should be kept in mind that these
have both hypercharge and $SU(2)$ couplings so that the analysis for this
reaction is not as
simple.

For the parameter choice in Case~2,
$m_{H_{\ell}}=93$ GeV and is very near the limit for observability
at LEP2.
We see from Fig.~10 that $\te_R\te_R$
is the dominant SUSY particle cross section over most of the range of
$P_L(e^-)$. $\tnu_{\ell}\tnu_{\ell}$ production also occurs, but since
$\tnu_{\ell}\to\tz_1\nu_{\ell}$, this process is invisible. $\te_R\te_L$
also occurs at a large rate and can
provide an opportunity for $\te_L$ mass measurement ($\sigma({\te_L\te_L})$
is an order of magnitude smaller).
The cross sections for $\tell_R\tell_R$ ($\tell_L\tell_L$) production shown
in the figure are summed over $\tmu$ and $\ttau$, and are smaller than
the corresponding selectron production cross sections because there is
no $t$-channel contribution in their case.
Finally, we note that 
$\tell_R\to \ell\tz_1$ and $\tell_L\to \ell\tz_1$ are the only allowed two body
decays of sleptons, so that the analysis of slepton production which
is signalled by like flavour, opposite sign acollinear lepton pairs, 
is free from complications from cascade decays.

Observation of a signal in the acollinear $e^+e^-+\esl$ or $\mu^+\mu^-+\esl$
channels without an accompanying signal in the $e^{\pm}\mu^{\mp}$ channel
would suggest a slepton hypothesis.
Focussing, for the moment on the dimuon channel, we can select out
$\tmu_R\tmu_R$
production by adjusting the beam polarization. 
Operating with $P_L(e^-)=-0.9$
reduces the $WW\to\mu\mu$ background to tiny levels, while enhancing 
production of right sleptons\cite{jlc}. Turning to prospects for the
smuon mass measurement,
we show in 
Fig.~11{\it a} the $E_{\mu}$ distribution in dimuon events
after the dilepton cuts of Sec. 2, 
for an integrated luminosity of 20 fb$^{-1}$.
The background level is denoted
by the dashed histogram while the arrows denote the theoretical 
endpoints. In Ref. \cite{jlc}, it was shown a mass measurement of $m_{\tmu_R}$
and $m_{\tz_1}$ can be made to $\sim \pm 1\%$, so we do not repeat the $\chi^2$
analysis here.

In Fig.~11{\it b}, we show the $E_{e^+}$ distribution for the same luminosity and
polarization. This distribution consists of two components: the solid
histogram is from $\te_R\te_R$ production, while the dashed histogram is from
$\te_R^-\te_L^+$ production. (Note: with $P_L(e^-)=-0.9$,
$\te_R^-\te_L^+$ is produced at a much higher rate than $\te_R^+\te_L^-$.)
Since $m_{\te_L}=238$ GeV here, in contrast to $m_{\te_R}=157$~GeV,
the 
endpoints for the two components differ significantly. Now, because 
we already know
$m_{\tz_1}$ from Fig.~11a, the upper endpoint will yield $m_{\te_L}$ to a 
similar degree of precision (a one-parameter fit may be made).
The energy distribution of $e^-$ is shown in Fig.~11{\it c}, 
again for both
$\te_R\te_R$ and $\te_R\te_L$ components (notice the different scale
from Fig.~11{\it b}).
The endpoints of $\te_R\te_L$ 
are completely contained within the $\te_R\te_R$ distribution. A clean sample
of $\te_R\te_L$ events can be isolated by requiring $E_{e^+}>75$ GeV 
from Fig.~11{\it b}.
Then the subsequent $E_{e^-}$ distribution can be plotted in 
Fig.~11{\it d}. These endpoints now depend on $m_{\te_L}$, $m_{\te_R}$ and 
$m_{\tz_1}$, so that consistency with the previous mass measurements 
as well as with the assumed production mechanisms, can be checked.

The other opportunity, in this case, is to identify $\tz_1\tz_2$ production
and measure $m_{\tz_2}$.
In this case, $\tz_2\to\ell\tell_L\to \ell\ell\tz_1$ with a branching
fraction of about
5\%, so one can search
again for acollinear dilepton pairs. SM backgrounds mainly
come from $WW$ production and we have to discriminate
the $\tz_2$ signal from $\tell\tell$ production
processes. We focus here on the dimuon signature, so we can ignore background
from $\te_L\te_R$ production. In this case, the $\tmu_R$ and $\tz_1$
masses are already well measured. 
We run with $P_L(e^-)=+0.9$, for a 50 fb$^{-1}$
data sample. To eliminate the $WW$ background, we require $\msl >250$ GeV, 
and $\Delta\phi (\mu\mu )<90^o$. In the remaining event sample, the muons
from $\tmu_L$ decay are very hard, so we require $E_{\mu}(fast)>75$ GeV,
and plot the $E_{\mu}(slow)$ distribution, which is shown in Fig.~12.
A small remaining background from $\tmu_L\tmu_L$ production
populates the 
$E_{\mu}(slow)>60$ GeV region, while the residual SM background populates
the $E_{\mu}(slow)<20$ GeV region. The $\tz_1\tz_2$ signal gives a distinct
upper endpoint, which depends on $\tz_1$, $\tz_2$ and $\tmu_L$ masses. 
By combining this information with the previous mass measurements, a constraint
on $m_{\tz_2}$ can be obtained.

\subsection{Case 3: Mixed chargino, slepton and sneutrino production}

In order to study the additional complications that arise when charginos
and sleptons are simultaneously accessible, we are led to consider Case~3
with
$(m_0,m_{1/2},A_0,\tan\beta ,sgn(\mu )=(200,100,0,2,-1)$, 
for which the spectrum of sparticles and production
cross sections are shown in Fig.~13. The polarization dependence of the
various cross sections is as in Fig.~6 and Fig.~10 and needs no 
further discussion.

Clearly, many
superpartners would be produced at NLC500. In this case,
$m_{H_\ell}=69$ GeV, so that the light Higgs boson 
would presumably have been discovered and studied at LEP2.
As can be seen from Fig.~13, $\tw_1$, $\tz_2$, $\tell_L$, $\tell_R$, $\tnu_L$
and possibly even $\tz_3$, $\tz_4$ and $\tw_2$ should be accessible. From
the cross sections shown in Fig.~13, we see that for
right polarized beams, $\te_R\te_R$ is the dominant process, while for
left polarized beams, in fact $\tnu_e\tnu_e$ is dominant, followed closely by
$\tw_1\tw_1$ production. In this case, the sneutrinos and sleptons are
significantly heavier than the charginos and neutralinos, so that
their cascade decays need to be incorporated in the analysis. For instance,
the
$\tnu$ with a mass of 207~GeV, decays visibly via
$\tnu\to\tz_2\nu$ (32\%), and $\tnu\to\tw_1\ell$ (61\%), with
$\tnu\to\tz_1\nu$ making up the balance.
The decay pattern of $\tell_L$ is similar although its direct decay to
the LSP occurs about 20\% of the time. In contrast, $\tell_R$ which
has no $SU(2)$ gauge interactions essentially
always decays via $\tell_R \to \ell\tz_1$. 
As can be seen 
from Fig.~13, there are so many SUSY processes taking place at 
$\sqrt{s}=500$ GeV that it is potentially 
difficult to isolate one from another!
The ability to tune the beam energy would certainly be 
desirable in this situation, so that one could sequentially study each
SUSY process as one passes production threshold. With
$m_{\tw_1}\simeq m_{\tz_2}\simeq 96$ GeV, however, at least 
$\tw_1\tw_1$, $\tz_1\tz_2$ and $\tz_2\tz_2$ would be occuring simultaneously,
even running NLC at energies as low as $\sqrt{s}\sim 300$ GeV. Incidently,
it is worth remarking that for this scenario, SUSY would 
certainly have been discovered\cite{fnal} 
at the Main Injector upgrade of the Tevatron,
and in several channels, so that it is again reasonable for us to focus our
attention on precision measurements of sparticle properties.

With unpolarized beams, one may attempt to measure $m_{\tw_1}$ and $m_{\tz_1}$
via the $\tw_1\tw_1\to\ell\nu\tz_1 q\bar{q}'\tz_1$ mode. We adopt the 
chargino cuts of Sec. 2, but find substantial background from $\tnu_L\tnu_L$
production which 
distorts the $E_{jj}$ distribution. In this case, running NLC at 
400 GeV, below $\tnu_L\tnu_L$ threshold allows a clean distribution of
$E_{jj}$ to be made. A mass measurement 
of $m_{\tw_1}$ and $m_{\tz_1}$ should be possible with a precision
of a few percent\cite{jlc}.

As a second measurement, we run with $P_L(e^-)=-0.9$ in an effort
to pick out
a $\te_R\te_R$ signal. We run with the slepton cuts of Sec. 2, 
for 20 fb$^{-1}$. The resulting distribution for SUSY signal as well as the 
SM and
other SUSY backgrounds is shown by the histograms in Fig.~14. 
The SUSY background comes
mainly from $\te_R\te_L$ production, and has a slightly different endpoint 
structure. A $m_{\te_R}$ and $m_{\tz_1}$ mass measurement to a few percent 
should be possible from this distribution, as shown in Ref. \cite{jlc}.

The large production cross section for $\tnu_e\tnu_e$ events for left polarized
beams can lead to some unique, almost background free signatures.
For instance, if one sneutrino decays via
$\tnu_e\to e\tw_1\to e+q\bar{q}'\tz_1$ and the other decays via
$\tnu_e\to e\tw_1\to e+\mu\nu\tz_1$, we obtain
spectacular $ee\mu +jets +\esl$ events.
For this topology, we require in addition $\eslt >25$ GeV and $\msl >40$ GeV.
We then plot the $E_e$ distribution, which should have endpoints determined by
$m_{\tnu_e}$ and $m_{\tw_1}$. We fit an appropriate function to the expected
$E_e$ distribution, and map out the $\chi^2$ values in the
$m_{\tnu_e}\ vs.\ m_{\tw_1}$ plane for a data set of 20 fb$^{-1}$ at
$P_L(e^-)=+0.9$. The minimum $\chi^2$ is shown in Fig.~15{\it a}, along with
contours of $\Delta\chi^2=2.3$ and 4.6 from the minimum. We obtain a measured
value of $m_{\tnu_e}=207.5\pm 2.5$ GeV and $m_{\tw_1}=96.9\pm 1.2$ GeV--
a 1\% measurement of these masses.
The $E_e$ distribution from data, along with the best fit,
are shown in Fig.~15{\it b}. Because our fit to the $E_e$ distribution
had been done for the same ``theory'' set, we have
double checked our procedure by repeating it for a nearby point
in parameter space. The results are
shown in Fig.~15{\it c} and 15~{\it d}; we see that we
obtain similar precision as in Fig.~15{\it a}.

The cascade decays of selectrons and smuons
can also lead to unique event signatures from  $\te_L\te_L$ or $\tmu_L\tmu_L$
production (of course, selectron production dominates).
One possibility is where each $\te_L\to e\tz_2\to e+\ell\bar{\ell}+\tz_1$,
which can give events with 6 leptons plus $\esl$. We focus only on this unusual
channel, although similar or perhaps
even better measurements can possibly be performed
in other channels. We require events with six leptons, no jets
and $\eslt >25$ GeV.
We further require the two fastest leptons to be of same flavor/opposite sign,
and then plot $E_{\ell}$ of the two hardest leptons. We study a 50 fb$^{-1}$
sample with $P_L(e^-)=+0.9$, and obtain a sample with $70$ events, all
from the signal (this implicitly assumes that the left selectrons and 
smuons are mass degenerate). The energy distribution of the two fastest leptons
has endpoints depending on $m_{\tell_L}$
and $m_{\tz_2}$. Again, by fitting this distribution, contours of
$\Delta\chi^2$ values are calculated, with the
result shown in Fig.~16{\it a}.
We see that a measurement of $m_{\tell_L}=221.6\pm 6$ GeV
and $m_{\tz_2}=94.7\pm 6.5$ GeV is possible.
The corresponding data and best fit
are shown below in Fig.~16{\it b}.
We note that the relationship
$m_{\tell_L}^2-m_{\tnu}^2=-M_W^2\cos 2\beta$
depends only on the $SU(2)$ gauge invariance
and so is very robust. A good measurement of $m_{\tell_L}$
and $m_{\tnu}$ can lead to a model-independent
determination of the parameter $\tan\beta$. Unfortunately for
the present case, $\tell_L$ and $\tnu$ are very close in mass and
a combination of these mass measurements only implies
$\tan\beta >1.8$. A better determination of $\tan\beta$ may be possible
for other values of parameters.

Finally, for Case~3, we examine $\tz_1\tz_2$ production. 
We require the dilepton
cuts of Sec. 2, but in addition require 2 opposite sign muons with 
$E_{vis.}<200$ GeV and $\phi (\ell\bar{\ell})<90^o$. We show the 
resulting dimuon 
invariant mass distribution in Fig.~17.
The solid histogram denotes the ``signal'' and  
the dashed one the background. In this 
case, the signal consists of 68\% $\tz_1\tz_2$ production, 21\% $\tnu\tnu$
production and 10\% $\tz_2\tz_2$ production. All these signal processes lead to
leptonically decaying $\tz_2$ plus missing energy. The $m(\mu^+\mu^-)$ plot
has a relatively sharp upper cutoff at $m_{\tz_2}-m_{\tz_1}$, allowing
an independent determination of $m_{\tz_2}$ given the information on $m_{\tz_1}$
from $\te_R$ or $\tw_1$ production processes discussed above.

\subsection{Case 4: Includes signals from top squark production}

Up to now, we have only considered cases with $A_0=0$. The weak scale
$A$-parameters, of course, do not vanish,
and are obtained by renormalization group evolution. The $A$-parameters
(and hence $A_0$) mainly affect the phenomenology of third generation
sfermions (and of gluinos, charginos and neutralinos via modifications
of their decay patterns\cite{THIRD})
which can be significantly lighter than their
first and second generation siblings. We are thus led to consider
Case~4 with
$(m_0,m_{1/2},A_0,\tan\beta ,sgn(\mu ))=(300,150,-600,2,+1)$, chosen
so as to lead to
a sparticle mass spectrum which contains a light top squark with
$m_{\tst_1}=180$ GeV which essentially always decays via $\tt_1\to b\tw_1$.
The SUSY particle masses and total cross sections as a
function of $P_L(e^-)$ are shown in Fig.~18.
In this case, $m_{H_{\ell}}=102$ GeV and
$m_{\tw_1}=110$ GeV, so both are just beyond the reach of LEP2.
The novel feature, not encountered previously, is the accessibility of
$\tt_1\tt_1$ production with a cross section of about 40~fb, independent
of $P_L(e^-)$. A stop with this mass is unlikely to be observable
even at the Main Injector upgrade of the Tevatron\cite{MRENNA}.

Charginos are the most copiously produced sparticles if the electron
beam is unpolarized or if its polarization is dominantly left-handed.
We first consider the prospects for measuring $\tw_1$ and $\tz_1$ masses
by analyzing
$\tw_1\tw_1$ production, using $P_L(e^-)=+0.9$. The difference from
the chargino mass analysis in Case~1 is that
now $\tw_1$ decays into 3-body $q\bar{q'}\tz_1$
and $\ell\nu\tz_1$ final states (with essentially the same branching fractions
as for $W$ decay) so that Eq. (3.1) is not applicable.
To isolate the $\tw_1\tw_1$ signal from SM background,
we use the $1\ell +$2-jets cuts of Sec.~II, and in addition require
$\msl > 240$~GeV.
The resulting distribution in $E_{jj}$ is shown in Fig.~19{\it a}.
The lower end point of this distribution depends on the jet
algorithm, while the upper end point is given by Eq. (3.1) 
with $m_4=m_{jj}=0$.
The mass analysis is thus not as straightforward as in the earlier cases.
In this case, we first consider the
scatter plot of $E_{jj}\ vs.\ m_{jj}$ shown in Fig.~19{\it b}.
For each value of $m_{jj}$, Eq.~(3.1) implies
a definite range of $E_{jj}$. The resulting
envelope of the ($m_{jj}$ dependent) endpoints of the $E_{jj}$ distribution
is shown as the solid contour in the Figure.
The chargino signal lies almost entirely within
this envelope, while some remaining background events
populate the outer regions. For an integrated luminosity of 50~fb$^{-1}$
the event sample is large enough  
to artificially force 2-body
kinematics onto the $\tw_1$ decay by requiring that the dijet
invariant mass $m_{jj}$ lie within some specified bin. In our analysis, we 
take four bins each of width 4 GeV, centered at $m_{jj}=22,\ 26,\ 30$ and 
$34$ GeV.
For each bin, the $E_{jj}$ distribution follows the form 
for $\tw_1\to\tz_1 W^*$, where $M_{W^*}$ is given by one of the previous
bin central values. 
The procedure for fitting $m_{\tw_1}$ and $m_{\tz_1}$ for the
forced two-body kinematics is the same as in previous cases, except that now 
we fit simultaneously to four different mass bins; the resulting
contours of
$\Delta\chi^2 =\chi^2-\chi^2_{min}$ are shown in Fig.~20{\it a} for a
data set of 50~fb$^{-1}$. We
find that $m_{\tw_1}=110.6\pm 5$ GeV (a 5\% measurement),
compared to the input value of $m_{\tw_1}=109.8$ GeV. Likewise,
$m_{\tz_1}=57.5\pm 2.5$ GeV, compared with the input of $m_{\tz_1}=57$ GeV.
These measurements may be improved by trying different bin choices or
different fitting procedures.
The $E_{jj}$ distribution and best fit to the data for the $m_{jj}$ bin 
centered at 30 GeV are shown in Fig.~20{\it b}. 
In this mass bin, the function
\begin{eqnarray*}
F(E) & = & N (1+exp[ (E_{min}+21.5-E)/3.7/\sigma_{E_{min}}] )^{-1} \\
& & \times (1+exp[ (-E_{max}+24+E)/1.9/\sigma_{E_{max}}] )^{-1}
\end{eqnarray*}
(where all energy parameters are in GeV) provides a fit to
the theoretical expectation for the $E_{jj}$ distribution.

Next, we focus on $\tz_2\tz_2$ events with a $P_L(e^-)=+0.9$ beam. The $\tz_2$
here decays via the three-body mode, with $B(\tz_2\to e\bar{e}\tz_1) =4.5\%$.
To obtain a clean event sample, we require four isolated leptons with no jets
in each event, and require as well
$\eslt >25$ GeV and 20 GeV$<E_{vis.}<400$ GeV. We look at only
$e^+e^-\mu^+\mu^-$ events, and veto any events with
$m(\ell^+\ell^-)=M_Z\pm 10$ GeV. We are left with just 9 signal events for
$\int{\cal L}dt=50$ fb$^{-1}$, with no SM background. The distribution of
the two like-flavour dilepton masses in each event
is shown in Fig.~21. This dilepton invariant mass is 
restricted to lie between 0 and $m_{\tz_2}-m_{\tz_1}$, so that the highest
mass value gives a lower bound on this mass difference (the fact that
we have a small number of events only means that there is a significant
chance that we may not find an event in the highest bin).
For 
$m_{\tz_2}-m_{\tz_1}=54$ GeV, we can combine results with the $m_{\tz_1}$
measurement from $\tw_1\tw_1$ events to deduce that 
$m_{\tz_2}\simeq 109\pm 3.5$ GeV
(where we have neglected the mismeasurement uncertainty on $m(\ell^+\ell^-)$).

Finally, we turn our attention to top-squarks.
In this case, $\tt_1 \to b\tw_1$, so that stop pair production
is signalled by events with two $b$-jets together with additional
jets or leptons from the decay products of the charginos and $\eslt$.
Since the $\tst_1\tst_1$
cross section varies hardly at all with $P_L(e^-)$, we run with
right polarized beams ($P_L(e^-)=-0.9$) to minimize backgrounds from $WW$
production. We search for events with $\ge 5$ jets, with two tagged
as $b$'s and no isolated leptons. We exclude hadronically decaying
$\tau$s by vetoing jets with one or three charged
particles as discussed in Sec.~1. This veto capability is crucial
to reduce large backgrounds from top quark production, where
one of the tops decays hadronically and the other decays via $t\to
b\tau\nu_{\tau}$. We also require
$\msl >140$ GeV. For a 50 fb$^{-1}$ sample of data,
we are left with a SUSY signal of 286 events, compared with
SM background of 36 events. At this point, we can plot the
energy distribution of the $b$-jets,
which should have endpoints determined by $m_{\tst_1}$ and $m_{\tw_1}$.
Again, we fit a function depending on these masses to a large sample of
generated top-squark pair events, and then obtain contours of $\Delta\chi^2$
for 50 fb$^{-1}$ of data. The results are plotted in Fig.~22{\it a}
while the resulting $E_b$ distribution of signal
plus background is shown in Fig.~22{\it b}, along with the best fit histogram.
We
see that a stop mass measurement of $m_{\tst_1}=182\pm 12$ GeV is obtained,
while $m_{\tw_1}=113.6\pm 8$ GeV. This measurement of $m_{\tw_1}$ is independent
of the measurement from $\tw_1\tw_1$ production described above and can
provide checks of the inferred sources of the signals. 
Since $m_{\tw_1}$ is determined to greater precision via the $\tw_1\tw_1$
channel, 
the precision of the stop mass measurement can be improved
by combining the two chargino mass measurements.
Since we had used this parameter point also to obtain the theoretical
fit to the $E_b$ distribution, we have repeated this exercise
for a somewhat different input for the $\tst_1$ mass in 
Fig.~21{\it c} and Fig.~21{\it d}. We see that a similar precision is obtained.


\section{Concluding Remarks}

Electron-positron collisions provide not only a clean facility for
the discovery of supersymmetric particles but also provide a unique locale
for detailed determination of their
properties \cite{lep2,peskin,nlc1,nlc,nlc3,grivaz,becker,feng,jlc}. 
The discovery
of charged sparticles is possible almost all the way up to the kinematic
limit for their production unless the mass difference between the parent
particle and the LSP becomes very small, so that the visible decay products
become very soft. Their main disadvantages, relative to hadron colliders,
are the lower centre of mass energy and generally smaller cross sections
so that considerable luminosity is needed for physics. These are balanced
by the clean experimental environment, simplicity of the initial
state and, at future linear colliders, the availability of
longitudinally polarized beams. The differences
make for complementary capabilities of hadron and $e^+e^-$ colliders.

In this paper, we have extended the pioneering work of Tsukamoto
{\it et. al.}\cite{jlc} and analysed what data from experiments at NLC500 might
look like if supersymmetry manifests itself via the minimal SUGRA framework,
and sparticles are kinematically accessible.
In Sec.~II, we have mapped out the discovery reach
of NLC500 in the $m_0-m_{1/2}$ plane for several sets of other parameters.
We have incorporated all the cascade decays
of the sparticles into our analyses. For the most part, the portion of
the SUGRA parameter space that can be probed at the NLC is determined by
where $\te_R$ and $\tw_1$ signals are observable, though there is
a small additional region that might be probed via signals from
$e^+e^- \to \tz_1\tz_2$ production. While the availability of longitudinally
polarized beams is extremely useful for reducing SM backgrounds (mainly from
$WW$ production) as well as for isolating various sparticle reactions, we
found that the beam polarization does not significantly
increase the SUSY reach if we
use the 5$\sigma$ level as our criterion for observability.

The combined reach of NLC500 (assuming an integrated
luminosity of 20~fb$^{-1}$), in the $m_0-m_{1/2}$ plane for $A_0=0$,
$\tan\beta=2$ and 10 and both signs of $\mu$ is shown by the lower
solid curve in Fig.~23{\it a-d}. These reach contours are not
particularly sensitive to the polarizability of the electron beam.
The dashed and dashed-dotted curves, respectively, denote the kinematic
limits for producing $\te_R$ and $\tw_1$ at Linear Colliders for
three different choices of $\sqrt{s}$. The small region where the solid
curve extends beyond the ``kinematic boundaries'' is the region that
should be explored via the $\tz_1\tz_2$ channel.
Finally, the upper solid curve
denotes the reach of the LHC via the $\eslt$ and multilepton channels
(the single lepton channel yields the greatest reach over
the whole space illustrated in Fig.~23)
as obtained in Ref.\cite{lhc}. We immediately see that {\it as far as,
and only as far as,} the supersymmetry reach is concerned, the LHC reach
would be comparable to that of a linear collider operating at $\sqrt{s}\sim
1500$~GeV. Of course, at this energy SM backgrounds from $2\to 3$ and $2 \to 4$
(vector boson scattering) processes would be important and need to be analysed
before drawing definite conclusions. 
We should mention though that NLC500 is guaranteed to find
the lightest neutral Higgs boson of the minimal model, and so can
exclude this framework if no signal is found. In contrast, the discovery
of the Higgs boson at the LHC is very difficult, and will 
certainly require machine
and detector performance at their design levels\cite{atlas}. 

It might also be worth noting that sparticles cannot be much beyond the
weak scale if SUSY is the new physics that stabilizes the symmetry breaking
sector of the SM. Several authors\cite{barbieri,castano}
have attempted to quantify this and
obtained {\it upper} bounds on sparticle masses; {\it e.g.} Anderson
and Casta\~no\cite{castano} have argued that the most favoured region from this
point of view is where $m_{1/2}\leq 200$~GeV, $m_0 \leq 200-300$~GeV.
Interestingly, the lightest neutralino is an acceptable mixed dark matter
candidate if SUGRA parameters are in this range\cite{brhlik}.
While this region
can partially be probed even at upgrades of the Tevatron\cite{fnal}, and
certainly at NLC500, it is worth keeping in mind that fine-tuning
considerations are qualitative, 
while the cosmological constraints
can be simply evaded, for instance, by allowing a small amount of $R$-parity
violation which could have no impact for collider searches. Thus,
the larger reach of the LHC or energy upgrades of the NLC 
provide a safety margin and
may prove
essential for a conclusive exploration of SUSY. Nonetheless,
the capability of NLC500 for discovering the  Higgs boson(s), in itself,
appears to us sufficient motivation for its construction.

But the Higgs sector aside, there are many other measurements that
are possible at Linear Colliders that would be very difficult 
or impossible at the LHC.
These include the precision measurement of sparticle masses, spins
and couplings which can lead to incisive tests of the underlying
framework\cite{jlc}. In Sec.~III, we have performed four case studies
to assess the prospects for measuring various sparticle masses. Our
study extends the earlier analysis in that we allow for all accessible
sparticle reactions, and attempt to devise strategies to measure
slepton masses even when these do not decay directly to the LSP. Also,
we demonstrate, for the first time, that at least for favourable
values of parameters, experiments at NLC500, with an integrated
luminosity of 20-50~fb$^{-1}$, should be able to
obtain masses of sneutrinos, scalar top quarks and even the second
lightest neutralino with a precision of a few percent. In this
analysis, the availability of a polarized electron
beam with 95\% longitudinal polarization has been assumed.
We have also identified strategies (that again 
make critical use of the polarization
capability of the NLC) to isolate various SUSY reactions from one another.
This separation facilitates the mass measurements, and also allows
for the measurement of SUSY cross sections which would be a first
step in the determination of sparticle couplings.
While some mass
measurements\cite{mass} are indeed possible at the LHC, the systematic
precision spectroscopy of sparticles appears to be possible only at
Linear Colliders.

As emphasized in Ref.\cite{jlc}, the measurement of sparticle masses
will allow stringent tests of the assumptions underlying the SUGRA
framework, and so provide a window to the symmetries of physics at
ultra-high scales. It could be that these assumptions will ultimately
prove to be incorrect. For instance, other models where
SUSY is broken at a relatively low scale $\sim 10-100$~TeV \cite{dine}
have been proposed. These models can have significantly different
mass patterns and can lead to very different phenomenology\cite{thomas}
from what we have considered. Sparticle spectroscopy will provide
guidance about the mechanism of supersymmetry breaking.
On a more pragmatic note, information about chargino and neutralino 
masses and couplings
obtained from experiments at NLC500 may prove very useful in
analysing the complicated cascade decay chains that should be present
in the LHC data sample. Indeed if it appears that NLC500 
is due to become operational significantly
after the LHC, we would advocate archiving the LHC data in 
a form suitable for subsequent reanalysis in light of new knowledge
gained from the NLC.
 
In summary, we have affirmed that if sparticles are kinematically
accessible at NLC500, it will not only be possible to detect the signals,
but it will also be possible to measure their masses with a precision
of 1-5\%, even if they do not directly decay into the LSP. While
electron beam polarization is not essential for SUSY discovery, it is
a crucial tool\cite{jlc} for these precision measurements. The complementary
capabilities\cite{peskin2} 
of NLC500 and the LHC cannot be overemphasized\cite{comp}. While the
LHC essentially probes the complete parameter space of the minimal
SUGRA framework (mainly via signals from the cascade decays of gluinos
and squarks), the reach of NLC500 (which indeed probes much of the
theoretically favoured region) is somewhat smaller. Higgs boson
searches, on the other hand, are much simpler at the NLC. Also, precision
measurements of sparticle masses and couplings, which are only possible
at the NLC, would be very helpful in disentangling LHC signals from heavy
squarks and gluinos (which may not even be accessible at NLC500). Working in
tandem, experiments at these facilities may allow us to uncover the 
mechanism of electroweak symmetry breaking, and perhaps also provide
clues about the symmetries of physics at ultra-high energy scales.


\acknowledgments

We thank H. Murayama, U. Nauenberg, F. Paige and M. Peskin for discussions,
and members of the NLC SUSY working group for discussions and motivation
to produce this report. XT is grateful to the High Energy Physics
Group at Florida State University for hospitality during his sabbatical
when this study was initiated.
This research was supported in part by the U.~S. Department of Energy
under grant numbers DE-FG-05-87ER40319 and DE-FG-03-94ER40833.
\bigskip
\appendix
\centerline{\bf Appendix: Production Cross-sections with Polarized Beams}
\bigskip

We present in this appendix various SM and SUSY cross sections 
retaining information on the polarization of the incoming beams. 
The notation used is that of Ref. \cite{bbkmt}.
First, we present lowest order SM cross sections for R or L polarized
incoming electrons and positrons. For SM fermion and gauge boson
pair-production, we have:

\begin{eqnarray*}
\frac{d\sigma}{dz}(e_{R\atop L} \bar{e}_{L\atop R} \to f \bar f)& = & 
\frac{N_f}{4 \pi} \frac{p}{E} \Phi_{f {R\atop L}}(z)
\end{eqnarray*}
where $z = \cos \theta$, ($\theta$ is the 
angle between incoming and outgoing fermions) 
$p$ and $E$ are the momentum magnitude and energy of 
either 
final state particle, $f= \mu,\ \tau,\ \nu_{\mu},\ \nu_{\tau}$, and $q$, and:

\begin{eqnarray*}
\Phi_{f {R\atop L}}(z)& = & 
e^4 \left[ \frac{q^2_f}{s^2} \left( E^2(1+z^2)+ m^2_f(1-z^2) \right)\right. \\
& & + \frac{( \alpha_e \pm \beta_e)^2}{(s-M^2_Z)^2 +M^2_Z \Gamma^2_Z}
\left( \left[ ( \alpha^2_f+ \beta^2_f)(E^2+p^2z^2)\pm\right.\right. \\
& &\left.\left. 4\alpha_f\beta_f Epz+(\alpha_f^2-\beta_f^2)m_f^2\right]\right) \\
& & \left. - \frac{2( \alpha_e \pm \beta_e)(s-M^2_Z)q_f}{s[(s-M^2_Z)^2 
+M^2_Z \Gamma^2_Z]}
\left( \alpha_f \left[ E^2(1+z^2)+m^2_f(1-z^2) \right] \pm 2 \beta_f Epz
\right) \right]
\end{eqnarray*}

\begin{eqnarray*}
\frac{d\sigma}{dz}(e_{R\atop L} \bar{e}_{L\atop R} \to Z Z)& = & 
\frac{e^4(\alpha_e \pm
\beta_e)^4 p}{4 \pi s \sqrt s} \left[ \frac{u(z)}{t(z)} + \frac{t(z)}{u(z)}
\right. \\
& & \left. + \frac{4 M^2_Z s}{u(z) t(z)}-M^4_Z \left( \frac{1}{t^2(z)} +
\frac{1}{u^2(z)} \right) \right]
\end{eqnarray*}
where $s$, $t(z)$, and $u(z)$ are the Mandelstam variables.

\begin{eqnarray*}
\frac{d\sigma}{dz}(e_{R\atop L} \bar{e}_{L\atop R} \to W^+ W^-)& = & 
\frac{e^4 p}{16 \pi s
\sqrt s \sin^4 \theta_W} \Phi_{WW {R\atop L}}(z)
\end{eqnarray*}
where:
\begin{eqnarray*}
\Phi_{WW R}(z)& = & \frac{4(\alpha_e + \beta_e)^2\tan^2\theta_W |D_Z|^2}{s^2}
\left[ U_T(z) (p^2 s + 3 M^4_W) + 4M^2_W p^2 s^2 \right]
\end{eqnarray*}
and
\begin{eqnarray*}
\Phi_{WW L}(z)& = & \frac{U_T(z)}{s^2} \left[ 3 + 2 (\alpha_e - \beta_e) \tan
\theta_W (s - 6 M^2_W) Re D_Z \right. \\
& & \left. + 4 (\alpha_e - \beta_e)^2 \tan^2 \theta_W (p^2 s + 3 M^4_W) |D_Z|^2
\right] + \frac{U_T(z)}{t^2(z)}+\\
& & + 8 (\alpha_e - \beta_e) \tan \theta_W M^2_W Re D_Z + 16 (\alpha_e -
\beta_e)^2 \tan^2 \theta_W M^2_W p^2 |D_Z|^2 \\
& & + 2 \left[ 1-2(\alpha_e - \beta_e) \tan \theta_W M^2_W Re D_Z \right]
\left[ \frac{U_T(z)}{s t(z)} - \frac{2 M^2_W}{t(z)} \right]
\end{eqnarray*}
where $U_T(z) = u(z) t(z) - M^4_W$, and $D_Z = (s - M^2_Z + i M_Z
\Gamma_Z)^{-1}$.

The expressions for lowest order MSSM Higgs Production include:

\begin{eqnarray*}
\frac{d\sigma}{dz}(e_{R\atop L} \bar{e}_{L\atop R} \to Z H_l)& = &
\frac{p}{16 \pi \sqrt s}
\frac{e^4 \sin^2 (\alpha +\beta)}{\sin^2 \theta_W \cos^2 \theta_W} 
\\ & & \times \frac{(\alpha_e \pm \beta_e)^2}{(s-M^2_Z)^2+M^2_Z \Gamma^2_Z}
(M^2_Z + E^2_Z -p^2 z^2)
\end{eqnarray*}
For $Z H_h$ production, replace $\sin^2(\alpha +\beta)$ with 
$\cos^2(\alpha +\beta)$.

\begin{eqnarray*}
\frac{d\sigma}{dz}(e_{R\atop L} \bar{e}_{L\atop R} \to H_l H_p)& = &
\frac{p^3}{16 \pi \sqrt s}
\frac{e^4 \cos^2 (\alpha +\beta)}{\sin^2 \theta_W \cos^2 \theta_W} 
\\ & & \times \frac{(\alpha_e \pm \beta_e)^2}{(s-M^2_Z)^2+M^2_Z \Gamma^2_Z} 
(1-z^2)
\end{eqnarray*}
For $H_h H_p$ production, replace $\cos^2(\alpha +\beta)$ with 
$\sin^2(\alpha +\beta)$.

\begin{eqnarray*}
\frac{d\sigma}{dz}(e_{R\atop L} \bar{e}_{L\atop R} \to H^+ H^-)& = &
\frac{e^4}{4 \pi} \frac{p^3}{\sqrt s} (1- z^2) \\
& & \times \left[ \frac{1}{s^2} + \left(\frac{2\sin^2 \theta_W -1}{2 \cos 
\theta_W \sin \theta_W}\right)^2 \frac{( \alpha_e \pm \beta_e)^2}{(s-M^2_Z)^2 
+ M^2_Z \Gamma^2_Z} \right. \\
& & \left. + \frac{1}{s} \left(\frac{2\sin^2 \theta_W -1}{\cos \theta_W 
\sin \theta_W}\right) \frac{( \alpha_e \pm \beta_e)(s-M^2_Z)}{(s-M^2_Z)^2 + 
M^2_Z \Gamma^2_Z}\right]
\end{eqnarray*}

For sfermion pair production ($\tilde f_i\bar{\tilde f_i}$, with 
$f=\mu ,\ \tau ,\ \nu_{\mu},\ \nu_{\tau},\ u,\ d,\ c,\ s,\ b$ 
and $i=L$ or $R$), we find:

\begin{eqnarray*}
\frac{d\sigma}{dz}(e_{R\atop L} \bar{e}_{L\atop R} \to \tilde f_i 
\bar{\tilde f_i})& = &
\frac{N_f}{256 \pi} \frac{p^3}{E^3} \Phi_{\tilde f_i {R\atop L}}(z)
\end{eqnarray*}
where
\begin{eqnarray*}
\Phi_{\tilde f_i {R\atop L}}(z)& = &  
e^4 (1-z^2) \left[ \frac{8 q_f^2}{s} + \frac{2 A^2_{f_i}
(\alpha_e \pm \beta_e)^2 s - 8 (\alpha_e \pm \beta_e) q_f A_{f_i} (s-M^2_Z)}
{(s-M^2_Z)^2 + M^2_Z \Gamma^2_Z} \right],
\end{eqnarray*}
where $A_{f_{L,R}}=2(\alpha_f \mp \beta_f )$. For the special case of
$\tst_1\bar{\tst_1}$ production, we have
$A_{t_1}=2(\alpha_f -\beta_f )\cos^2\theta_t 
+ 2(\alpha_f +\beta_f )\sin^2\theta_t$; for $\tst_2\bar{\tst_2}$ production, 
simply switch $\cos^2\theta_t$ with $\sin^2\theta_t$.
Also,
\begin{eqnarray*}
\frac{d\sigma}{dz}(e_{R\atop L}\bar{e}_{L\atop R} \to \tilde t_1 
\bar{\tilde t_2})& = &
\frac{48 \pi \alpha^2}{\sqrt s} \frac{(\alpha_e \pm \beta_e)^2 \beta^2_t 
\cos^2 \theta_t \sin^2 \theta_t} {[ (s-M^2_Z)^2 + M^2_Z \Gamma^2_Z]} p^3 (1-z^2)
\end{eqnarray*}

For selectron pair production, we find
\begin{eqnarray*}
\frac{d\sigma}{dz}(e_{R\atop L} \bar{e}_{L\atop R} 
\to \tilde e_L \bar{\tilde e_L})& = &
\frac{1}{256 \pi} \frac{p^3}{E^3} \Phi_{\tilde e_L {R\atop L}}(z)
\end{eqnarray*}
where
\begin{eqnarray*}
\Phi_{\tilde e_L R}(z)& = & \Phi_{\tilde \mu_L R}(z)
\end{eqnarray*}
and
\begin{eqnarray*}
\Phi_{\tilde e_L L}(z)& = & \Phi_{\tilde \mu_L L}(z) + \sum\limits_{i=1}^{4} 
\frac{|A^e_{\tilde Z_i}|^4 s(1-z^2)}{[2E(E-pz)-m^2_{\tilde e_L}+m^2_{\tilde
Z_i}]^2} \\
& & -8e^2 (1-z^2) \sum\limits_{i=1}^{4} \frac{|A^e_{\tilde Z_i}|^2}
{[2E(E-pz)-m^2_{\tilde e_L}+m^2_{\tilde Z_i}]} \\
& & \times \left[ 1 + \frac{(\alpha_e -
\beta_e)^2 s (s-M^2_Z)}{(s-M^2_Z)^2 + M^2_Z \Gamma^2_Z} \right] \\
& & + \sum\limits_{i<j=1}^{4} \frac{|A^e_{\tilde Z_i}|^2|A^e_{\tilde Z_j}|^2 s
(1-z^2)}{[2E(E-pz)-m^2_{\tilde e_L}+m^2_{\tilde Z_i}]
[2E(E-pz)-m^2_{\tilde e_L}+m^2_{\tilde Z_j}]}
\end{eqnarray*}
Similarly,
\begin{eqnarray*}
\frac{d\sigma}{dz}(e_{R\atop L} \bar{e}_{L\atop R} \to 
\tilde e_R \bar{\tilde e_R})& = &
\frac{1}{256 \pi} \frac{p^3}{E^3} \Phi_{\tilde e_R {R\atop L}}(z) .
\end{eqnarray*}
where
\begin{eqnarray*}
\Phi_{\tilde e_R L}(z)& = & \Phi_{\tilde \mu_R L}(z)
\end{eqnarray*}
and $\Phi_{\tilde e_L L}(z) \to \Phi_{\tilde e_R R}(z)$ with the substitutions:
$A^e_{\tilde Z_i} \to B^e_{\tilde Z_i}$, $m_{\tilde e_L} \to m_{\tilde e_R}$,
and $( \alpha_e - \beta_e ) \to ( \alpha_e + \beta_e )$.
Also, 
\begin{eqnarray*}
\frac{d\sigma}{dz}(e_R\bar{e}_L \to \tilde e_L \bar{\tilde e_R} )& = &
\frac{d\sigma}{dz}(e_L\bar{e}_R \to \tilde e_R\bar{\tilde e_L} )= 0,
\end{eqnarray*}
while
\begin{eqnarray*}
\frac{d\sigma}{dz}(e_L\bar{e}_L \to \tilde e_L\bar{\tilde e_R}  )& = &
\frac{1}{32 \pi s} \frac{p}{E} \left[ \sum\limits_{i=1}^{4}
\frac{|A^e_{\tilde Z_i}|^2 |B^e_{\tilde Z_i}|^2 m^2_{\tilde Z_i} }
{[E_{\te_L}-pz+a_{\tilde Z_i}]^2} \right. \\
& & \left. + \sum\limits_{i<j=1}^{4} \frac{ 2 m_{\tilde Z_i} m_{\tilde Z_j} Re(
A^e_{\tilde Z_i} A^{e \ast}_{\tilde Z_j} B^{e \ast}_{\tilde Z_i} 
B^e_{\tilde Z_j}) }
{[E_{\te_L}-pz+a_{\tilde Z_i}] [E_{\te_L}-pz+a_{\tilde Z_j}] } \right]
\end{eqnarray*}
where $a_{\tz_i}=\frac{m_{\tz_i}^2-m_{\te_L}^2}{2E}$. Also,
\begin{eqnarray*}
\frac{d\sigma}{dz}(e_R\bar{e}_R \to \tilde e_R\bar{\tilde e_L}  )& = &
\frac{1}{32 \pi s} \frac{p}{E} \left[ \sum\limits_{i=1}^{4}
\frac{|A^e_{\tilde Z_i}|^2 |B^e_{\tilde Z_i}|^2 m^2_{\tilde Z_i} }
{[E_{\te_R}-pz+a_{\tilde Z_i}]^2} \right. \\
& & \left. + \sum\limits_{i<j=1}^{4} \frac{ 2 m_{\tilde Z_i} m_{\tilde Z_j} Re(
A^e_{\tilde Z_i} A^{e \ast}_{\tilde Z_j} B^{e \ast}_{\tilde Z_i} 
B^e_{\tilde Z_j}) }
{[E_{\te_R}-pz+a_{\tilde Z_i}] [E_{\te_R}-pz+a_{\tilde Z_j}] } \right]
\end{eqnarray*}
where now $a_{\tz_i}=\frac{m_{\tz_i}^2-m_{\te_R}^2}{2E}$.

For $\tnu_e$ pair production, we find
\begin{eqnarray*}
\frac{d\sigma}{dz}(e_R\bar{e}_L \to  \tilde \nu_e \bar{\tilde \nu_e})& = &
\frac{d\sigma}{dz}(e_R\bar{e}_L \to  \tilde \nu_{\mu} \bar{\tilde \nu_{\mu}})
\end{eqnarray*}
while
\begin{eqnarray*}
\frac{d\sigma}{dz}(e_L\bar{e}_R & \to & \tilde \nu_e \bar{\tilde \nu_e}) =
\frac{p^3 E}{8 \pi} (1-z^2) \\
& & \times \left[ \frac{4 e^4 (\alpha_{\nu} - \beta_{\nu})^2 (\alpha_e - 
\beta_e)^2 }{(s-M^2_Z)^2 + M^2_Z \Gamma^2_Z } +
\frac{g^4 \sin^4 \gamma_R}{[2E(E-pz)+ m^2_{\tw} -m^2_{\tilde \nu_e}]^2}\right.\\
& & + \frac{g^4 \cos^4 \gamma_R}{[2E(E-pz)+ m^2_{\tw_2} -m^2_{\tilde \nu_e}]^2} 
- \frac{4 e^2 g^2 (\alpha_{\nu} - \beta_{\nu}) (\alpha_e - \beta_e) (s-
M^2_Z)\sin^2\gamma_R}{[(s-M^2_Z)^2 + M^2_Z \Gamma^2_Z]
[2E(E-pz)+ m^2_{\tw} -m^2_{\tilde \nu_e}]}  \\
& & - \frac{4 e^2 g^2 (\alpha_{\nu} - \beta_{\nu}) 
(\alpha_e - \beta_e) (s-M^2_Z)\cos^2\gamma_R}
{[(s-M^2_Z)^2 + M^2_Z \Gamma^2_Z] 
[2E(E-pz)+ m^2_{\tw_2} -m^2_{\tilde \nu_e}]}\\
& & +\left. \frac{2g^4\sin^2\gamma_R\cos^2\gamma_R}
{[2E(E-pz)+ m^2_{\tw} -m^2_{\tilde \nu_e}]
[2E(E-pz)+ m^2_{\tw_2} -m^2_{\tilde \nu_e}]}\right] .
\end{eqnarray*}

For neutralino pair production, we find
\begin{eqnarray*}
\frac{d\sigma}{dz}(e_{R\atop L}\bar{e}_{L\atop R} \to \tilde Z_i \tilde Z_j)& = &
\frac{k}{8 \pi s \sqrt s} \left( M_{\tilde e \tilde e {R\atop L}} + 
M_{Z Z{R\atop L}} + M_{Z \tilde e {R\atop L}} \right)
\end{eqnarray*}
where
\begin{eqnarray*}
M_{\tilde e \tilde e R} & = & 2 |B^e_{\tilde Z_i}|^2 |B^e_{\tilde Z_j}|^2 G_t(
m_{\tilde Z_i},m_{\tilde Z_j},m_{\tilde e_R},z)
\end{eqnarray*}
\begin{eqnarray*}
M_{\tilde e \tilde e L} & = & 2 |A^e_{\tilde Z_i}|^2 |A^e_{\tilde Z_j}|^2 G_t(
m_{\tilde Z_i},m_{\tilde Z_j},m_{\tilde e_L},z)
\end{eqnarray*}
\begin{eqnarray*}
M_{Z Z {R\atop L}} & = & 
\frac{4 e^2 |W_{ij}|^2 (\alpha_e \pm \beta_e)^2} {(s-M^2_Z)^2
+ M^2_Z \Gamma^2_Z} \left[ s^2 - (m^2_{\tilde Z_i}-m^2_{\tilde Z_j})^2 \right.
\\ & & \left. - 4 (-1)^{\theta_i + \theta_j} s m^2_{\tilde Z_i} 
m^2_{\tilde Z_j} + 4 s k^2 z^2 \right]
\end{eqnarray*}
\begin{eqnarray*}
M_{Z \tilde e R} & = & \frac{-e(-1)^{(\theta_i +\theta_j+1)} 
(\alpha_e + \beta_e )(s-M^2_Z)} {2[(s-M^2_Z)^2 +
M^2_Z \Gamma^2_Z}] \\
& & \times \left[ Re(W_{ij} B^{e \ast}_{\tilde Z_i}
B^e_{\tilde Z_j}) G_{st} (m_{\tilde Z_i},m_{\tilde Z_j},m_{\tilde e_R},z)
\right. \\
& & \left. + (-1)^{(\theta_i +\theta_j)} 
Re(W_{ij} B^e_{\tilde Z_i} B^{e \ast}_{\tilde Z_j}) G_{st} (
m_{\tilde Z_i},m_{\tilde Z_j},m_{\tilde e_R},-z) \right]
\end{eqnarray*}
and
\begin{eqnarray*}
M_{Z \tilde e L} & = & \frac{-e (\alpha_e - \beta_e )(s-M^2_Z)} 
{2[(s-M^2_Z)^2 +M^2_Z \Gamma^2_Z]} \\
& & \times \left[ Re(W_{ij} A^{e \ast}_{\tilde Z_i}
A^e_{\tilde Z_j}) G_{st} (m_{\tilde Z_i},m_{\tilde Z_j},m_{\tilde e_L},z)
\right. \\
& & \left. + (-1)^{\theta_i + \theta_j} Re(W_{ij} A^e_{\tilde Z_i} 
A^{e \ast}_{\tilde Z_j}) G_{st} ( m_{\tilde Z_i},m_{\tilde Z_j},m_{\tilde e_L},
-z) \right] .
\end{eqnarray*}
The functions $G_t$ and $G_{st}$ are defined in Ref. \cite{bbkmt}.

For chargino pairs, we have
\begin{eqnarray*}
\frac{d\sigma}{dz}(e_L\bar{e}_R \to \tw_1 \bar{\tw_1})& = &
\frac{1}{64 \pi s} \frac{p}{E} \left( M_{\gamma \gamma L} + M_{ZZL} 
+ M_{\gamma ZL} \right. \\
& & \left. + M_{\tilde \nu \tilde \nu L} + M_{\gamma \tilde \nu L} + 
M_{Z \tilde \nu L} \right)
\end{eqnarray*}
and
\begin{eqnarray*}
\frac{d\sigma}{dz}(e_R\bar{e}_L \to \tw_1 \bar{\tw_1})& = &
\frac{1}{64 \pi s} \frac{p}{E} \left( M_{\gamma \gamma R} + M_{ZZR} + 
M_{\gamma ZR} \right)
\end{eqnarray*}
where
\begin{eqnarray*}
M_{\gamma \gamma L} & = & M_{\gamma \gamma R}= \frac{16 e^4}{s} \left[ 
E^2(1+z^2 ) + m^2_{\tw_1} (1-z^2 ) \right]
\end{eqnarray*}
\begin{eqnarray*}
M_{ZZ{R\atop L}} & = & \frac{16 e^4 \cot^2 \theta_W s}
{(s-M^2_Z)^2+M_Z^2 \Gamma^2_Z} \left[ 
(x^2_c +y^2_c )(\alpha_e \pm \beta_e )^2 [E^2 (1+z^2 )+m^2_{\tw_1} 
(1-z^2 )] \right. \\
& & \left. - 2 y^2_c (\alpha_e \pm \beta_e )^2 m^2_{\tw_1} \mp 4 x_c y_c
(\alpha_e \pm \beta_e )^2 Epz \right]
\end{eqnarray*}
\begin{eqnarray*}
M_{\gamma Z{R\atop L}} & = & \frac{-32 e^4 (\alpha_e \pm \beta_e ) \cot \theta_W 
(s-M^2_Z )}{(s-M^2_Z)^2 +M^2_Z \Gamma^2_Z} \\
& & \left[ x_c [E^2 (1+z^2 )+m^2_{\tw_1} (1-z^2 )] \mp 2 y_c Epz \right]
\end{eqnarray*}
\begin{eqnarray*}
M_{\tilde \nu \tilde \nu L} & = & \frac{2 e^4 \sin^4 \gamma_R}{\sin^4 \theta_W} 
\frac{s (E-pz)^2}{[E^2 + p^2 -2Epz + m^2_{\tilde \nu}]^2}
\end{eqnarray*}
\begin{eqnarray*}
M_{\gamma \tilde \nu L} & = & \frac{-8 e^4 \sin^2 \gamma_R}{\sin^2 \theta_W} 
\frac{[(E-pz)^2 + m^2_{\tw_1}]}{[E^2 + p^2 -2Epz + m^2_{\tilde \nu}]}
\end{eqnarray*}
and
\begin{eqnarray*}
M_{Z \tilde \nu L} & = & \frac{8 e^4 (\alpha_e - \beta_e ) \cot \theta_W \sin^2 
\gamma_R}{\sin^2 \theta_W} 
\frac{s(s-M^2_Z )}{(s-M^2_Z)^2 +M^2_Z \Gamma^2_Z } \\
& & \times \left[ \frac{(x_c -y_c )[(E-pz)^2 + m^2_{\tw_1} ] +2y_c
m^2_{\tw_1}}{E^2 + p^2 -2Epz + m^2_{\tilde \nu}} \right] .
\end{eqnarray*}
For $\tw_2 \bar{\tw_2}$ production, replace $x_c$ with $x_s$,
$y_c$ with $y_s$, $\sin \gamma_R$ with $\cos \gamma_R$ and $m_{\tw_1}$ 
with $m_{\tw_2}$.
Finally,
\begin{eqnarray*}
\frac{d\sigma}{dz}(e_L\bar{e}_R \to \tw_1 \bar{\tw_2})& = &
\frac{e^4}{64 \pi} \frac{p}{E} \left[ M_{ZZL} +M_{\tilde \nu \tilde \nu L}+
M_{Z \tilde \nu L} \right]
\end{eqnarray*}
and
\begin{eqnarray*}
\frac{d\sigma}{dz}(e_R\bar{e}_L \to \tw_1 \bar{\tw_2})& = &
\frac{e^4}{64 \pi} \frac{p}{E} M_{ZZR}
\end{eqnarray*}
where
\begin{eqnarray*}
M_{ZZ{R\atop L}} & = & \frac{4 (\alpha_e \pm \beta_e )^2 (\cot \theta_W + \tan \theta_W
)^2 }{(s-M^2_Z )^2 + M^2_Z \Gamma^2_Z } \left[ (x^2 +y^2 )(E^2 +p^2 z^2
\right. \\
& & \left. - \Delta^2 - \xi m_{\tw_1} m_{\tw_2} )  
+ 2 x^2 \xi m_{\tw_1} m_{\tw_2} \mp 4xyEpz \right]
\end{eqnarray*}
\begin{eqnarray*}
M_{\tilde \nu \tilde \nu L} & = & \frac{2 \sin^2 \gamma_R \cos^2 \gamma_R}{\sin^4
\theta_W} \frac{[(E-pz)^2 - \Delta^2 ]}{[2E(E- \Delta)-2Epz+ m^2_{\tilde \nu}
-m^2_{\tw_1}]^2}
\end{eqnarray*}
\begin{eqnarray*}
M_{Z \tilde \nu L} & = & \frac{-4 \theta_y (\alpha_e - \beta_e )(\cot \theta_W 
+ \tan \theta_W ) \sin \gamma_R \cos \gamma_R (s-M^2_Z )}{\sin^2 \theta_W [(s-
M^2_Z)^2 + M^2_Z \Gamma^2_Z ]} \\
& & \times \frac{(x-y)[(E-pz)^2 - \Delta^2 - \xi m_{\tw_1} 
m_{\tw_2} ]+2x \xi m_{\tw_1} m_{\tw_2} }{[2E(E- \Delta)-2Epz+ 
m^2_{\tilde \nu} -m^2_{\tw_1}]}.
\end{eqnarray*}

The final cross section can be calculated from
\begin{eqnarray*}
\sigma =f_L(e^-)f_L(e^+)\sigma_{LL}+f_L(e^-)f_R(e^+)\sigma_{LR}+
f_R(e^-)f_L(e^+)\sigma_{RL}+f_R(e^-)f_R(e^+)\sigma_{RR},
\end{eqnarray*}
where $f_L$ and $f_R$ are defined in Sec. I, and $\sigma_{ij}$ ($i,j=L,R$) 
refers to the cross section from $e_i^- e_j^+$ annihilation.
The above formulae have been incorporated into the event generator
ISAJET\cite{isajet}.

%

%
\newpage


\begin{figure}
\caption[]{Regions of the $m_0\ vs.\ m_{1/2}$ plane where various
sparticle pair production reactions are kinematically accessible.
For $\tz_i\tz_j$ pairs, instead we plot the 10 fb cross section
contours. We take $A_0=0$, $\tan\beta =2$ and $m_t=180$ GeV.
In {\it a}), we take $\mu <0$, while in {\it b}) we take $\mu >0$.
The regions denoted
by TH are excluded by theoretical constraints, while the region labelled EX
is excluded by experimental constraints.
}
\end{figure}
\begin{figure}
\caption[]{Same as Fig. 1, except for $\tan\beta =10$.
}
\end{figure}
\begin{figure}
\caption[]{Various lowest order SM cross sections (in fb) for NLC at 
$\sqrt{s}=500$ GeV,
as a function of $e^-$ polarization parameter $P_L(e^-)$.
}
\end{figure}
\begin{figure}
\caption[]{Regions of the $m_0\ vs.\ m_{1/2}$ plane where
selectrons, charginos and neutralinos are accessible at the $5\sigma$
level above SM backgrounds, after various cuts discussed in the text.
The $\tz_1\tz_2$ contour denotes the boundary of the added region where
$\geq 10$ $b\bar{b}+\esl$ events (10 fb cross section) should be obtained with
$P_L(e^-)=0.9$
above negligible background. 
We have taken $\sqrt{s}=500$ GeV and have 
assumed 20 fb$^{-1}$ of integrated luminosity.
The SUGRA parameters taken are the same as for Fig. 1.
In frame {\it a}), we show the reach for both 
polarized and unpolarized beams, and in addition, the location of case study
points 1, 2 and 3.
}
\end{figure}
\begin{figure}
\caption[]{Same as Fig. 4, except $\tan\beta =10$.
}
\end{figure}
\begin{figure}
\caption[]{Various lowest order SUSY cross sections (in fb) for NLC at 
$\sqrt{s}=500$ GeV,
as a function of $e^-$ polarization parameter $P_L(e^-)$, for case study
point \#1. We also show on the left the masses of only the accessible 
superpartners and Higgs bosons.
}
\end{figure}
\begin{figure}
\caption[]{{\it a}) Histograms of events after cuts versus missing mass 
$\msl$ for chargino pair events and also for SM and SUSY backgrounds.
In {\it b}), we plot the $E_{jj}$ distribution after requiring 
$\msl >240$ GeV; the arrows denote the kinematic endpoints which are
related to the $\tw_1$ and $\tz_1$ masses.
}
\end{figure}
\begin{figure}
\caption[]{In {\it a}), we plot the $\msl$ distribution after cuts
to isolate the $\tz_1\tz_2\to b\bar{b}+\esl$ signal from SM and SUSY
backgrounds. The arrows denote the expected kinematic endpoints of the
signal distribution. After a cut of $\msl >300$ GeV, we plot $m_{jj}$
for the two detected $b$-jets; the distribution is related to the light
Higgs boson $H_{\ell}$ mass, denoted by the arrow.
In {\it c}), we plot the $E_{jj}$ distribution from the two $b$-jets;
the kinematic endpoints, which depend on $m_{\tz_2}$, $m_{\tz_1}$
and on $m_{H_{\ell}}$, are denoted by the tips of the arrows.
}
\end{figure}
\begin{figure}
\caption[]{In {\it a}), comparison of data to a theory fit depending on
$m_{\tz_1}$ and $m_{\tz_2}$ yields the minimum $\chi^2$ value
shown, and also the $\Delta\chi^2 =2.3$ and 4.6 contours. In {\it b}),
a 50 fb$^{-1}$ `data' sample is shown as well as the best fit 
theory distribution, for the $\msl$ distribution for $\tz_1\tz_2$
search using case study point 1.
}
\end{figure}
\begin{figure}
\caption[]{Same as Fig. 6, except for case study point \#2.
}
\end{figure}
\begin{figure}
\caption[]{For 95\% right polarized $e^-$ beams, we show 
in {\it a}) the $E_{\mu}$ distribution after cuts
for case study point \#2. The endpoints, denoted by the tips
of the arrows, depend on $m_{\mu_R}$
and $m_{\tz_1}$. In {\it b}), we show the $E_{e^+}$ distribution
which has two main components from $\te_R\te_R$ production and from
$\te_R\te_L$ production. The corresponding endpoints are shown as well.
In {\it c}), the $E_{e^-}$ distribution has a different structure 
than that shown in frame {\it b}). Finally, after requiring 
$E_{e^+}>75$ GeV in {\it d}), the $E_{e^-}$ distribution from
$\te_R\te_L$ is cleanly isolated as discussed in Sec.~IIIB of the text.
}
\end{figure}
\begin{figure}
\caption[]{The $E_{\mu}^{slow}$ distribution from $\tz_1\tz_2$
production (after requiring $E_{\mu}^{fast}>75$ GeV) shown along
with SM and SUSY backgrounds. The upper endpoint of the solid histogram
yields information on $m_{\tz_2}$, given knowledge of $m_{\tz_1}$
and $m_{\tmu_L}$.
}
\end{figure}
\begin{figure}
\caption[]{Same as Fig. 6, except for case study point \#3.
}
\end{figure}
\begin{figure}
\caption[]{Distribution in $E_{e^-}$ from $\te_R\te_R$ signal and
SM and SUSY backgrounds, for case study point \#3, using a right
polarized $e^-$ beam.
The endpoints yield information on $m_{\te_R}$ and $m_{\tz_1}$.
}
\end{figure}
\begin{figure}
\caption[]{The distribution of $E_e$ in $ee\mu +$jets$+\esl$ events from
$\tnu_e\tnu_e$ production yields information on $m_{\tnu_e}$ and $m_{\tw_1}$.
After comparing `data' to a theoretical fit, we plot the location of the
minimum $\chi^2$ value, and contours of $\Delta\chi^2=2.3$ and 4.6 in
frame {\it a}). In {\it b}), we show the data compared to the best theory 
fit. In {\it c}), we again plot $\Delta\chi^2$ contours for a nearby
point in parameter space. Similar precision to frame {\it a}) is attained.
In {\it d}), we show the $E_e$ distribution for data and best fit 
for the point taken in frame {\it c}).
}
\end{figure}
\begin{figure}
\caption[]{The $E_{\ell}$ distribution from the two hardest leptons in
$\te_L\te_L\to6\ell$ events leads to a mass measurement of $m_{\te_L}$
and $m_{\tz_2}$. In {\it a}), we plot the minimum $\chi^2$ and also
contours of $\Delta\chi^2 =2.3$ and 4.6. In {\it b}), we show a sample 
of `data' and also the best fit obtained from frame {\it a}).
}
\end{figure}
\begin{figure}
\caption[]{Distribution in dimuon invariant mass showing the
$\tz_1\tz_2\to\mu^+\mu^- +\esl$ contribution and SM and SUSY backgrounds.
The upper endpoint of the signal distribution is bounded by 
$m_{\tz_2}-m_{\tz_1}$, and leads to a measurement of $m_{\tz_2}$.
}
\end{figure}
\begin{figure}
\caption[]{Same as Fig. 6, except for case study point \#4.
}
\end{figure}
\begin{figure}
\caption[]{In frame {\it a}), we show the $E_{jj}$ distribution from
$\tw_1\tw_1\to \ell+jj+\esl$ events along with SM and SUSY backgrounds.
No distinctive endpoints are evident, due to the three-body chargino 
decay kinematics. In {\it b}), we show a scatter plot of 
$E_{jj}\ vs.\ m_{jj}$. The signal is kinematically constrained to lie
below the solid contour.
}
\end{figure}
\begin{figure}
\caption[]{
By requiring $m_{jj}$ to lie within 4 narrow mass bins, 
we force 2-body kinematics
onto the signal distribution. Then, for each bin, the endpoints in the $E_{jj}$
distribution depend on $m_{\tw_1}$ and $m_{\tz_1}$. In {\it a}), we plot
the minimum $\chi^2$ value, along with contours of $\Delta\chi^2$, after
performing a common fit to the four $m_{jj}$ bins mentioned in the text.
In frame {\it b}), we show the `data' sample from the $m_{jj}=30$ GeV bin, 
compared to the best fit theory distribution.
}
\end{figure}
\begin{figure}
\caption[]{$\tz_2\tz_2$ production leads to a clean sample of 
$e^+e^-\mu^+\mu^- +\esl$ events. A plot of $m_{\ell^+\ell^-}$
yields a distribution bounded by $m_{\tz_2}-m_{\tz_1}$, which gives
an estimate of $m_{\tz_2}$.
}
\end{figure}
\begin{figure}
\caption[]{$\tst_1\tst_1$ production leads to events containing
$\ge 5$ jets$+\esl$, where two of the jets are tagged as $b$-jets.
A plot of $E_b$ yields information on $m_{\tst_1}$ and $m_{\tw_1}$.
In frame {\it a}), we plot minimum $\chi^2$ and contours of
$\Delta\chi^2$ from a comparison of theory to data. In frame {\it b}),
we show the `data' sample, and the best fit theory distribution.
In {\it c}), a similar analysis is made for a nearby point in
parameter space, and similar precision is obtained. Frame {\it d})
shows the associated best fit to the `data' distribution.
}
\end{figure}
\begin{figure}
\caption[]{The $m_0\ vs.\ m_{1/2}$ plane is illustrated for $A_0=0$,
and for $\tan\beta =2$ and 10, and $\mu <0$ and $\mu >0$. The regions denoted
by TH and EX are excluded by theortical and experimental constraints,
respectively. The dashed lines show contours of $m_{\te_R}$ while the
dot-dashed lines are contours of $m_{\tw_1}$, and roughly denote
the kinematic reach of NLC500, NLC1000 and NLC1500 via selectron
and chargino pair production channels. The lower solid contour
shows the reach of NLC500 in parameter space as obtained using
our simulation. The upper solid
contour shows the reach of LHC with 10 fb$^{-1}$ of integrated 
luminosity, in the $1\ell +$jets$+\eslt$ channel\cite{lhc}.
}
\end{figure}


\begin{references}
\bibitem{comp} Our considerations here have been confined to the minimal
SUGRA framework with $R$-parity conservation. The complementarity
of the LHC and NLC extends well beyond this. For instance, if $R$-parity
is violated by baryon number non-conserving operators, the LSP (and, perhaps
also other sparticles) will have new hadronic decay modes. The $\eslt$
signature at the LHC will thus no longer be viable, and multilepton event
cross sections will be reduced. Although no detailed studies of these
scenarios exist either for the LHC or for NLC, it seems reasonable to
expect that new particle production at the NLC will be readily detectable
via an excess of spherical events, while at the LHC, the signal will
be harder to detect (particularly if sparticles other than the LHC also
decay via $R$-violating operators with significant branching fractions). 
On the other hand, it should be remembered that we do not have any real
theory of SUSY breaking parameters, and it could be that, fine-tuning 
considerations notwithstanding, none of the
sparticles may be accessible at NLC500 or even NLC1000.
\end{references}
\end{document}